\documentclass[floats,floatfix,showpacs,amssymb,prd,twocolumn,superscriptaddress,nofootinbib]{revtex4-1}

\usepackage{anyfontsize}
\usepackage{amssymb}
\usepackage{amsmath}
\usepackage{amsfonts}
\usepackage{amsbsy}
\usepackage{newtxtext}
\usepackage{newtxmath}

\usepackage{subfigure}
\usepackage{gensymb}
\usepackage{soul}
\usepackage{wrapfig}

\def\namedlabel#1#2{\begingroup
    #2%
    \def\@currentlabel{#2}%
    \phantomsection\label{#1}\endgroup
}

\usepackage{mathrsfs}

\usepackage{bm}

\newcommand{\dG}{\Delta G/G_\text{N}}

\newcommand{\GN}{G_{\rm N}}

\usepackage[utf8]{inputenc}

\usepackage{graphicx}
\usepackage{epsfig}
\usepackage{epstopdf}

\usepackage[usenames,dvipsnames]{xcolor}

\usepackage{verbatim,mathtools,needspace,enumitem,etoolbox}

\definecolor{linkcolor}{rgb}{0.0,0.3,0.5}

\usepackage[unicode, colorlinks=true, linkcolor=linkcolor, citecolor=linkcolor, filecolor=linkcolor,urlcolor=linkcolor, pdfusetitle]{hyperref}

\usepackage{epstopdf}

\usepackage{bm}
\usepackage{dcolumn}

\usepackage{longtable}
 \usepackage[normalem]{ulem}

\definecolor{romared}{RGB}{142,0,28}
\hypersetup{colorlinks=true,
            citecolor=romared,
            linkcolor=romared,
            urlcolor=romared}
\begin{document}

\title{A local resolution of the Hubble tension: \\The impact of screened fifth forces on the cosmic distance ladder}

\author{Harry Desmond}
\email{harry.desmond@physics.ox.ac.uk}
\affiliation{Astrophysics, University of Oxford, Denys Wilkinson Building, Keble Road, Oxford OX1 3RH, UK}

\author{Bhuvnesh Jain}
\email{bjain@physics.upenn.edu}
\affiliation{Center for Particle Cosmology,
Department of Physics and Astronomy,
University of Pennsylvania,
209 S. 33rd St., Philadelphia, PA 19104, USA}

\author{Jeremy Sakstein}
\email{sakstein@physics.upenn.edu}
\affiliation{Center for Particle Cosmology,
Department of Physics and Astronomy,
University of Pennsylvania,
209 S. 33rd St., Philadelphia, PA 19104, USA}

\raggedbottom

\begin{abstract}
The discrepancy between the values of the Hubble constant $H_0$ derived from the local distance ladder and the cosmic microwave background provides a tantalising hint of new physics. We explore a potential resolution involving screened fifth forces in the local Universe, which alter the Cepheid calibration of supernova distances. In particular, if the Cepheids with direct distance measurements from parallax or water masers are screened but a significant fraction of those in other galaxies are not, neglecting the difference between their underlying period--luminosity relations biases the local $H_0$ measurement high. This difference derives from a reduction in the Cepheid pulsation period and possible increase in luminosity under a fifth force. We quantify the internal and environmental gravitational properties of the Riess et al. distance ladder galaxies to assess their degrees of screening under a range of phenomenological models, and propagate this information into the $H_0$ posterior as a function of fifth force strength. We consider well-studied screening models in scalar--tensor gravity theories such as chameleon, K-mouflage and Vainshtein, along with a recently-proposed mechanism based on baryon--dark matter interactions in which screening is governed by local dark matter density. We find that a fifth force strength $\sim5-30\%$ that of gravity can alleviate (though not resolve) the $H_0$ tension in some scenarios, around the sensitivity level at which tests based on other distance ladder data can constrain this strength. Many of our models reduce the tension below $3\sigma$, but reduction below $2\sigma$ is unlikely possible while maintaining self-consistency of the distance ladder. Although our analysis is exploratory and based on screening models not necessarily realised in full theories, our results demonstrate that new physics-based \emph{local} resolutions of the $H_0$ tension are possible, supplementing those already known in the pre-recombination era.
\end{abstract}

\date{\today}

\maketitle

\section{Introduction}

The Hubble parameter $H_0$, describing the expansion rate of the Universe today, is a critical quantity in cosmology. Perhaps the most significant discrepancy that currently exists in the standard cosmological model, $\Lambda$CDM, is that when measured locally (out to $\sim$100 Mpc) $H_0$ is found to be $\sim$74 km s$^{-1}$ Mpc$^{-1}$, while high-redshift observations imply $\sim$67 km s$^{-1}$ Mpc$^{-1}$. The current most precise estimates from these two regimes are $74.03 \pm 1.42$ km s$^{-1}$ Mpc$^{-1}$ from the local distance ladder \cite{R16,R18,R19} (hereafter R16, R18 \& R19) and $67.4 \pm 0.5$ km s$^{-1}$ Mpc$^{-1}$ from the cosmic microwave background (CMB) \cite{planck} (hereafter \textit{Planck}), a tension of $4.4 \sigma$. These are supplemented at low redshift by independent analyses such as H0LiCOW, which uses time delay distances to low-redshift lensed quasars to derive $H_0 = 73.3^{+1.7}_{-1.8}$ km s$^{-1}$ Mpc$^{-1}$ \cite{Wong}, and at high redshift by galaxy and Ly$\alpha$ measurements of baryon acoustic oscillations calibrated against elemental abundances from Big Bang nucleosynthesis \cite{Addison} and the ``inverse distance ladder'' \cite{IDL}, yielding $H_0 =  66.98 \pm 1.18$ km s$^{-1}$ Mpc$^{-1}$ and $H_0 =  67.77 \pm 1.30$ km s$^{-1}$ Mpc$^{-1}$ respectively.

Prospective resolutions to this disagreement fall into one of three classes. The first is to modify the statistical framework to refine the measure of the tension (e.g. \cite{Cardona, Zhang, Feeney, Bennett}). While a range of estimates for the probability of distance ladder and CMB concordance are derivable from plausible methodologies, the magnitude of the raw discrepancy (derived simply by combining errorbars in quadrature) has been growing, and is now sufficiently large that almost all statistical approaches suggest strong disagreement. The second is to argue that one or more of the $H_0$ pipelines suffer from systematics which have not been fully accounted for in the fiducial analyses. This is more plausible for the distance ladder measurements than the CMB due to the complexity of the calibrations between rungs of the ladder and the range of models required to describe the astrophysical objects involved, although some fraction of the discrepancy could be due to systematics in CMB measurements (e.g. \cite{Spergel}). For the distance ladder, re-analyses have focused on the treatment of outliers in the Cepheid period--luminosity relation (PLR) \cite{Efstathiou, Becker}, the dependence of supernova luminosity on the mass and star formation rate of the host galaxy \cite{Rigault}, the parallaxes of Milky Way Cepheids \cite{Benedict, Shanks} and the geometric distances to the anchor galaxies \cite{Fitzpatrick, Humphreys}.

\begin{figure*}[ht]
    \centering
    \includegraphics[width=0.475\textwidth]{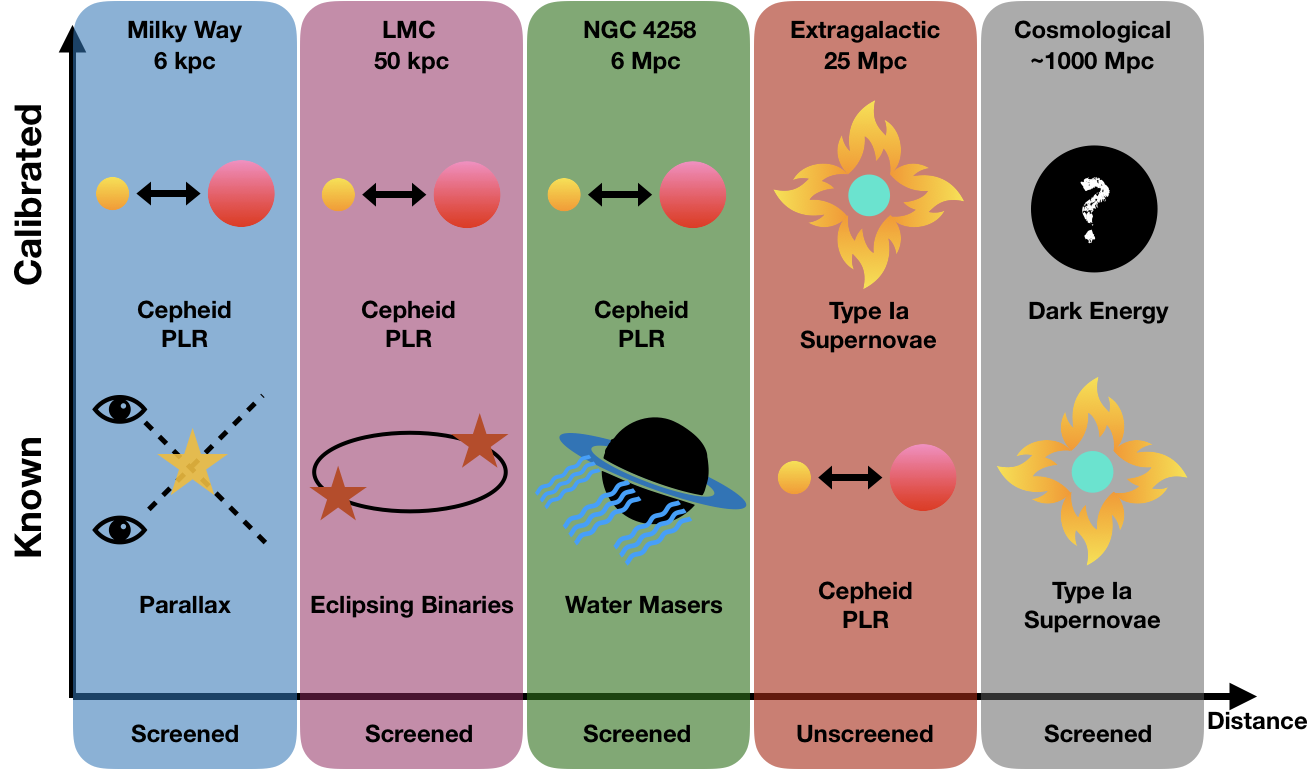}\hfill
    \includegraphics[width=0.475\textwidth]{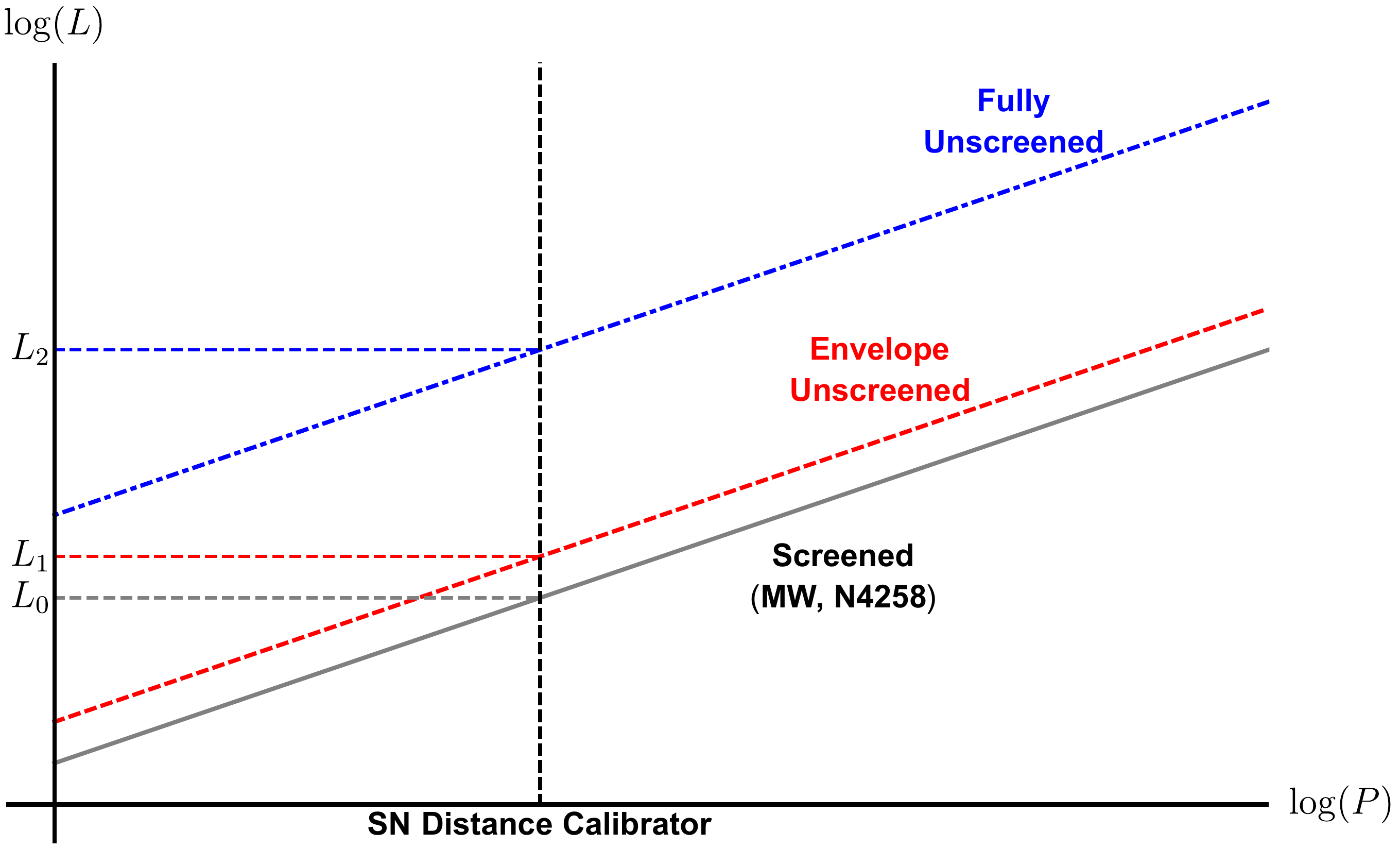}
    \caption{\emph{Left:} The distance ladder. Each vertical segment represents a rung, as indicated at the very top. Distances to objects in the upper part are calibrated by means of the lower indicator. A fiducial screening status of each rung is shown under the distance axis: a rung is labelled as unscreened when that is the case for at least one indicator. In this work we consider SNe to be screened, although this may not be the case in some models; we consider this further in Appendix \ref{Appendix:SN}. \emph{Right:} A schematic representation of the Cepheid period--luminosity relation (PLR) when various parts of a Cepheid are unscreened. The grey solid line shows the Newtonian relation as traced out by the screened Cepheids in the MW and N4258. The red dashed line shows the relation for Cepheids with unscreened envelopes and the blue dotdashed line for both envelope and core unscreened. The distances between the lines indicate that unscreening the core has a larger effect than unscreening the envelope. 
    Assuming an unscreened Cepheid lies on the Newtonian PLR causes its luminosity and hence distance to be underestimated, as shown by the vertical line at fixed measured period. This causes the inferred $H_0$ to be biased high.\newline}
    \label{fig:stairway_to_heaven}
\end{figure*}

The final class of solutions to the $H_0$ tension is to postulate new physics that introduces a difference between the expansion rate inferred locally and at recombination. Although requiring a deviation from $\Lambda$CDM, and therefore less conventional, this could naturally explain the fact that a variety of local measurements cluster around $73$ km s$^{-1}$ Mpc$^{-1}$ and high-redshift ones around $67$ km s$^{-1}$ Mpc$^{-1}$. For example, it has been proposed that the Milky Way is near the centre of a statistically underdense region of size potentially several hundred Mpc, which would generate streaming velocities away from us and hence bias local measurements of $H_0$ high \cite{Wojtak, Kenworthy}. The CMB measurements are also sensitive to the physics of recombination, which may be modified to alter the $H_0$ constraint \cite{Lin:2018nxe,Chiang:2018xpn}. Alternatively, an increase in the number of relativistic degrees of freedom $N_\text{eff}$ by $\sim$0.4--1 (``dark radiation'') could partially resolve the tension \cite{Wyman, Aubourg, Hinshaw, planck_1}; this is degenerate with a change in the effective strength of gravity between the epochs of Big Bang Nucleosynthesis and recombination \cite{Steigman}. There is also a significant degeneracy with the systematic lensing power parameter $A_L$ \cite{planck_1}, whose measured deviation from unity is not understood. A component of dark energy at very high redshift \cite{Karwal,Poulin:2018cxd,Kaloper:2019lpl}, an interaction of dark energy \cite{DE_interaction,Agrawal:2019dlm}, an injection of energy from scalar fields at recombination \cite{Randall} or the decay of dark matter into dark radiation at the time of matter-radiation equality \cite{Berezhiani:2015yta,Bringmann:2018jpr,Pandey:2019plg} or later \cite{Vattis:2019efj} may also push the CMB value upwards. A full high-redshift solution would appear to require an exotic physics scenario however, as none of the commonly-considered extensions to $\Lambda$CDM offer an entirely satisfactory solution \cite{planck_1, Bernal, Alam}.

We propose a local resolution to the $H_0$ tension relying on theoretically and observationally well-motivated physics beyond the standard model. Specifically, we describe how a partially screened fifth force naturally leads to an overestimate of $H_0$ in distance-ladder analyses if the gravitational properties of Cepheids calibrating the period--luminosity relation (principally in the Milky Way and N4258) differ from those of Cepheids in galaxies with Type Ia supernovae (hereafter simply ``SNe''), which are used to extend the distance ladder to higher redshift. The emergence of a fifth force is a generic prediction of theories that couple new dynamical degrees of freedom to matter, postulate new interactions between objects, or seek to provide a dynamical explanation of dark energy \cite{Clifton:2011jh,Koyama:2015vza,Burrage:2017qrf,Sakstein:2018fwz} (although see \cite{Heckman:2019dsj} for an exception). The phenomenon of screening---a reduction of fifth force strength in regions of strong gravitational field---was originally proposed to keep scalar--tensor theories of gravity consistent with tests of the equivalence principle, the inverse square law and post-Newtonian tests of gravity within the Solar System while allowing them to impact astrophysical and cosmological observables, and has now been recognised as a fairly generic property of theories with fifth forces \cite[see e.g.][and references therein]{Khoury_rev, Jain_rev, Joyce_rev, Burrage:2016bwy, Burrage:2017qrf, Sakstein:2018fwz,Ishak:2018his}. The Universe's accelerated expansion has been hypothesised to arise from new fields in this way \cite{Copeland:2006wr,Clifton:2011jh}, and the low-energy limits of UV complete theories are expected to contain screened fifth forces on a range of scales \cite{Davis, Jain,Vikram,Koyama:2015oma, Sakstein:2015zoa,Sakstein:2015aac, Sakstein:2016lyj, Babichev:2016jom, Sakstein:2016oel, Sakstein:2016ggl, Sakstein:2017bws, Sakstein:2017pqi,Adhikari:2018izo,Khoury:2018vdv}. Empirical searches for these are growing in sophistication \cite[e.g.][]{Burrage:2017qrf,Burrage:2019yle,Brax:2018iyo}, with several indications that they may in fact be of use in accounting for certain astrophysical phenomena \cite{Lombriser, Burrage, Desmond_PRD, Desmond_warp, Naik}.

In this paper we develop a framework for propagating gravitational enhancements due to fifth forces into the distance constraints on extragalactic Cepheid hosts, and hence $H_0$ when SN measurements are included. We implement a range of phenomenological screening models within this framework---in which the degree of screening is set by either environmental or intra-halo gravitational variables---in order to quantify the bias expected in $H_0$ under various scenarios for fifth force behaviour. Some of our screening proxies correspond to commonly studied mechanisms including chameleon \cite{Khoury:2003aq,Khoury04}, symmetron \cite{Symmetron}, dilaton \cite{Brax}, K-mouflage \cite{kinetic} and Vainshtein \cite{Vainshtein} screening. In a companion paper~\cite{rho_DM}, we present a new screening mechanism in which interactions between dark matter and baryons make Newton's constant a function of local dark matter density: this enables us to add $\rho_\text{DM}$ to our list of possible screening proxies. This theory is particularly interesting because the same dark matter--baryon interactions could drive cosmic acceleration \cite{Berezhiani:2016dne}. Other proxies that we consider do not (yet) correspond to known screening models but simply quantify different aspects of gravitational environment.

The structure of the paper is as follows. In Sec.~\ref{sec:distance_ladder} we provide more information on the determination of $H_0$ by the distance ladder method, the origin and nature of screened fifth forces, and the impact of fifth forces on Cepheids. In Appendix \ref{Appendix:SN} we consider the further effect of unscreening SNe. In Sec.~\ref{sec:method} we describe our models for setting the fifth force strengths experienced by the objects in the distance ladder dataset, then use this information to rederive the galaxies' distances and hence the $H_0$ constraint. In Sec.~\ref{sec:results} we formulate and carry out consistency tests within the distance ladder data that limit the possible action of a fifth force, then use these bounds to derive maximal viable modifications to $H_0$ by our mechanism. Sec.~\ref{sec:disc} provides caveats and possible systematics, speculates on future theoretical and observational developments, and discusses the broader ramifications of our work for expansion rate inference and the study of fifth forces generally. Sec.~\ref{sec:conc} concludes.

\section{Unscreening the distance ladder}
\label{sec:distance_ladder}

\subsection{Overview}
\label{sec:heuristic}

Determining $H_0$ via the distance ladder involves calibrating distances to a number of different types of objects in the local Universe in order to reach cosmological redshifts. Within the Milky Way (MW), parallaxes of Cepheids are measured with \textit{Hipparcos}, \textit{HST} and \textit{Gaia}, allowing direct determination of their distance. This enables their apparent magnitudes to be converted into luminosities, which, along with the measured pulsation periods, provides an absolute calibration of the PLR. This is believed to be near-universal across the Cepheid population \cite{Garcia-Varela:2013xda}, although there is a weak dependence on metallicity \cite{Storm}. Further constraining power for the PLR comes from Cepheids in the Large Magellanic Cloud (LMC) and M31 \cite{Persson, Sebo}, the distances to which are measured independently with detached eclipsing binaries \cite{Piet}, and in N4258 whose distance is calibrated with a water maser \cite{Humphreys}. The distance ladder is then extended with a sample of Cepheids in more distant galaxies. The periods and apparent magnitudes of these Cepheids are measurable, but their luminosities and hence distances must be inferred by situating them on the PLR. The galaxies are chosen to host SNe so that their Cepheid distances allow calculation of the SN absolute magnitudes. As SNe are standardizable candles, this information may be used to calibrate a cosmological SN sample and hence extend the distance--redshift relation beyond $z=1$. The slope of this relation as $z \rightarrow 0$ provides the value of $H_0$ \cite{R09, R11}.

A key assumption of this method is that the physics governing the objects used in the ladder (principally Cepheids and SNe) is homogeneous between rungs. If this assumption breaks down, a bias will be induced in the distance determinations, which will propagate into the cosmological SN analysis and hence the inferred value of $H_0$. In general, any novel physics affecting Cepheids, SNe, masers, or eclipsing binaries should be parameterised, implemented in the full likelihood function and marginalised over to rederive the $H_0$ posterior. 

In this work we explore a specific class of modifications to Cepheids involving the action of a screened fifth force. The effect of this on the distance ladder derives primarily from the impact of a new force on the pulsation and energy generation rate of Cepheids. Taking a scalar--tensor theory for illustration, and assuming a Compton wavelength for the scalar much larger than the size of the star (i.e. the field is relatively light), the fifth force scales with distance as $1/r^2$ and hence simply augments the gravitational force, effectively increasing the value of Newton's constant from $G_{\rm N}$ to $G > G_{\rm N}$.
We show below that this reduces the Cepheid pulsation period while increasing the luminosity. If all Cepheids experience the same effective $G$, this does not affect distance inferences as the PLR traced out by the calibration Cepheids in the MW and N4258 will coincide with that of the Cepheids in the SN hosts. However, if the calibration and cosmological Cepheids experience \emph{different} values of $G$ on average, the offset in PLR normalizations will bias the inferred Cepheid luminosities and hence their distances. A schematic of the distance ladder, and the sensitivity of its rungs to fifth forces, is shown in Fig.~\ref{fig:stairway_to_heaven} (left). 

The scenario we envision is that a difference in $G$ between the MW, N4258, and other galaxies comes about through a difference in their screening properties, caused in turn by a difference in their gravitational environments. In particular, if all MW and N4258 Cepheids are screened by their dense surroundings (as the Solar System must be to suppress the fifth force locally) and a significant fraction of the extragalactic Cepheids are not, the inferred luminosities of the latter will be lower than the true ones, and hence the distances to their hosts will be underestimated. For fixed redshift, this biases the estimate of $H_0$ high. Additionally, the uncertainty in the screening properties of the Cepheids would broaden the $H_0$ posterior. We therefore expect this scenario to reduce the tension of the distance ladder and CMB $H_0$ estimates. Sec.~\ref{sec:F5_cepheids} quantifies the effect of $G > G_\text{N}$ on Cepheid properties, while Sec.~\ref{sec:method} further propagates this information into distance estimates to the SN hosts, and hence $H_0$.

We consider a variety of phenomenological proxies for the degree of screening and use them to separate Cepheids into those sensitive or insensitive to the fifth force. These proxies are inspired by the chameleon \cite{Khoury04}, symmetron \cite{Symmetron}, dilaton \cite{Brax}, K-mouflage \cite{kinetic, Brax_K}, Vainshtein \cite{Vainshtein}, and baryon--dark matter interaction \cite{rho_DM} screening mechanisms, although they are typically more general and/or empirically motivated:

\begin{enumerate}

\item {\it Externally sourced Newtonian potential $\Phi$.}
This is the environmental contribution to screening by the thin-shell mechanisms chameleon, symmetron, and dilaton. This model has the additional degree of freedom of the Compton wavelength of the scalar field $\lambda_C$, which determines the radius out to which sources contribute to the screening potential. We explore the range $0.5 < \lambda_C/\text{Mpc} < 50$.\label{it:1}

\item {\it Externally sourced acceleration $a$.}
This is the environmental contribution to screening under a kinetic mechanism such as K-mouflage. Although kinetic models do not possess a mass or Compton wavelength, we introduce a cutoff scale $R_\text{max}$ beyond which sources are considered to have decoupled from the screening behaviour of the test object. This behaves similarly to $\lambda_C$ in the thin-shell models, and we consider the same range $0.5 < R_\text{max}/\text{Mpc} < 50$.\label{it:2}

\item {\it Externally sourced curvature.}
This is quantified by the Kretschmann scalar $K$ and is a proxy for screening in the Vainshtein mechanism. We use a cutoff scale $R_\text{max}$ as above: in massive gravity \cite{deRham:2014zqa,deRham:2016nuf} and massive Galileon \cite{deRham:2017imi,Sakstein:2018pfd} models this would again correspond to the Compton wavelength of the field.\label{it:3}

\item {\it Galaxy stellar mass.}
The galaxy's $V$-band luminosity $L_\text{gal}$ is used as a proxy for the galaxy's stellar mass, which provides the bulk of the baryonic contribution to self-screening.\label{it:4}

\item {\it Dynamical mass.}
This is inferred from the galaxies' neutral hydrogen gas, and in particular the HI linewidth measured at $20\%$ of peak flux, $W_{20}$. It is a proxy for self-screening.\label{it:5}

\item {\it Halo virial mass $M_\text{vir}$.}
Although not directly measurable, and hence subject to larger uncertainty than $L_\text{gal}$ or $W_{20}$, theoretically this quantifies more directly the screening properties of the halo as a whole.\label{it:6}

\item {\it Local dark matter density.}
Interactions between dark matter and baryons, in particular the dark energy model of \cite{Berezhiani:2016dne}, predict $G$ to be a function of the local dark matter density $\rho_\text{DM}$, with stronger modifications in less dense environments \cite{rho_DM}. This screening mechanism is phenomenologically distinct from the others by being sensitive to the position of the test object within its dark matter halo, and can therefore assign different degrees of screening to different objects within the same halo.\label{it:7} 

\end{enumerate}

\subsection{Fifth force effects on Cepheids}
\label{sec:F5_cepheids}

The luminosity of Cepheid stars is due to a thin hydrogen burning shell surrounding their inert helium core (there may be a small amount of core helium burning but the luminosity is due primarily to the shell). The pulsations on the other hand are driven entirely by the star's envelope. In particular, a layer of doubly-ionised helium acts as an energy dam because small contractions of the star result in further ionization rather than a temperature or pressure increase (which would counter the contraction) \cite{1980tsp..book.....C}. The pulsation occurs because this energy is eventually released after sufficient contraction. In \cite{rho_DM} we describe in detail how a fifth force impacts Cepheids, finding two main effects: first, if the envelope is unscreened then the dynamics driving the pulsation are modified, and second, if the core is unscreened the luminosity will be enhanced (a faster burning rate is required to balance the stronger gravity). For chameleon models and similar, the Cepheid core is screened and the envelope is unscreened \cite{Jain}. In the case where the fifth force is due to a dark matter--baryon interaction, as we discuss further below, the entire star is unscreened. Other screening mechanisms may exhibit either of these behaviours. We now derive the modification to the PLR due to both of these effects.

We begin with an unscreened envelope. The period of a Cepheid's pulsation is set heuristically by the free-fall time, $P \propto (G \rho)^{-1/2}$ \cite{1980tsp..book.....C}, although more accurate results with precisely the same scaling follow from the linear adiabatic wave equation \cite{Sakstein:2013pda}. We model the impact of the fifth force as an effective increase in Newton's constant $G$; for theories in which the fifth force is not of the same form as Newtonian gravity, $G$ can be thought of as an average of the radial fifth force profile weighted with a kernel that prioritises the regions most important for driving the pulsations \cite{Jain}. This means that the PLR of unscreened Cepheids is normalised lower in $P$ by a factor $(G/G_{\rm N})^{1/2}$ than that of the screened Cepheids in the MW and N4258\footnote{
The PLR is also calibrated using eclipsing binary distances to the LMC and M31; in Sec.~\ref{sec:disc} we discuss further our assumption that these also effectively calibrate the screened PLR.} that comprise the calibration sample:
\begin{equation}
P \propto (1+\Delta G/G_{\rm N})^{-1/2} \Rightarrow \Delta \log(P) = - \frac{1}{2} \log(1+\Delta G/G_{\rm N}),
\end{equation}
where $\Delta G \equiv G - G_{\rm N}$ and $\log$ has base 10 throughout. A reduction in period shifts the PLR equivalently to an increase in luminosity by an amount determined by the logarithmic slope of the PLR, which we call $A$:
\begin{equation}\label{eq:LchangeP}
\Delta \log(L) = \frac{A}{2} \log(1+\Delta G/G_{\rm N}),
\end{equation}
We take a fiducial value for $A$ of $1.3$ \cite{Becker}. Assuming the same PLR for both calibration and unscreened samples therefore causes the luminosity of the unscreened Cepheids to be underestimated. 

Next we calculate the change to the PLR if the core is unscreened. As described in detail in \cite{rho_DM}, we have modified the stellar structure code MESA \cite{Paxton:2010ji,Paxton:2013pj,Paxton:2015jva,Paxton:2017eie} to simulate stellar evolution for $G_{\rm N}\rightarrow G_{\rm N}(1+\Delta G/ G_{\rm N})$. An example is shown in Fig. \ref{fig:HR}, where we plot the Hertzsprung--Russell track for a 5$M_\odot$ Cepheid for different values of $\Delta G/G_{\rm N}$ as indicated in the legend. 
We see that the luminosity of the blue loop in the instability strip is enhanced for stronger gravity. To quantify this, we have run a grid of models in the mass range $5M_\odot\le M_\text{ceph} \le 13 M_\odot$ with $Z=0.0006$, the mean metallicity in the R16 sample. (Metallicity variations will add to the scatter, but are subdominant to the effects of changing $G$.) We find that $\Delta\log(L)$ at the blue edge of the instability strip, a convenient comparison point, is well fit by a linear relation:

\begin{figure}
    \centering
    \includegraphics[width=0.5\textwidth]{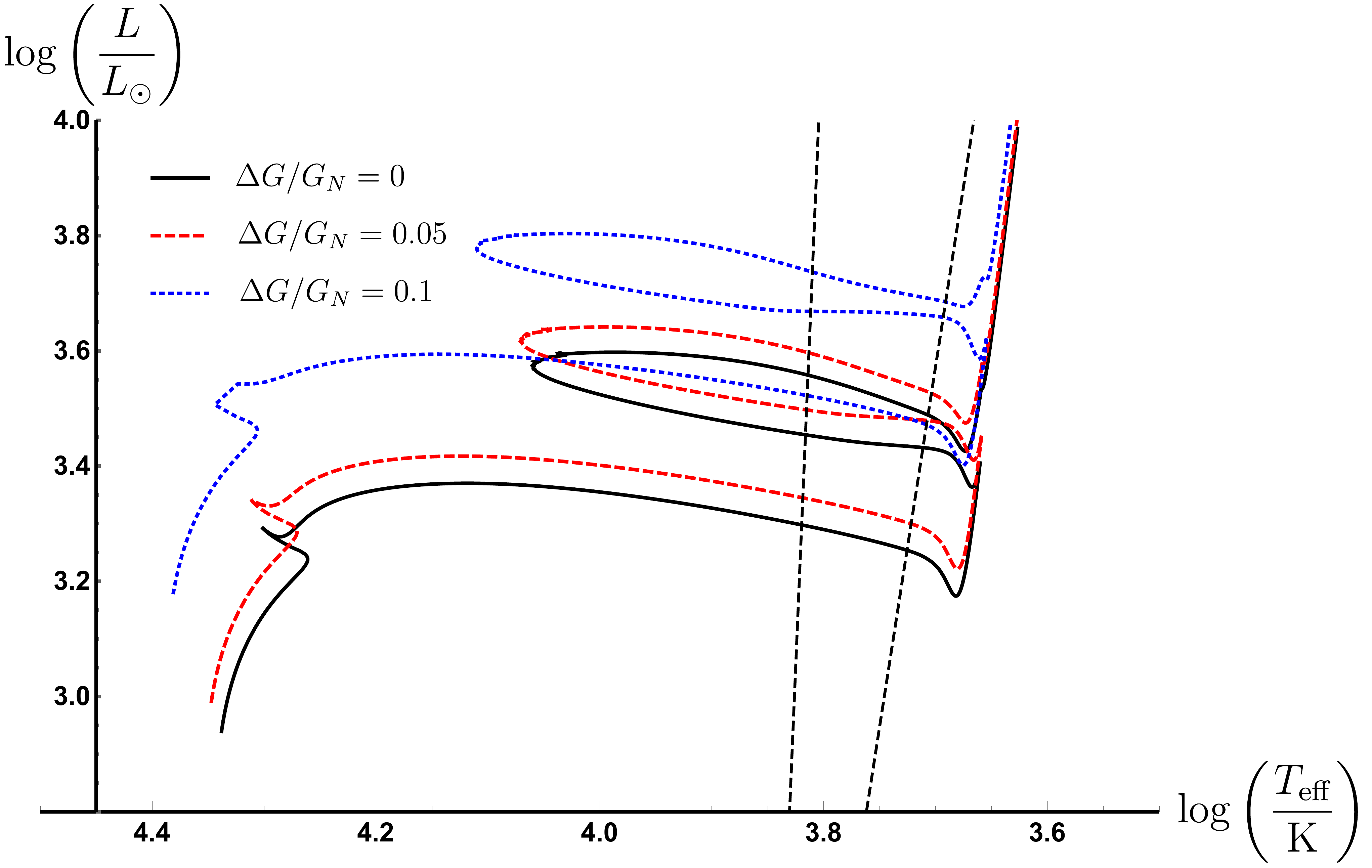}
    \caption{The Hertzsprung--Russell track for a 5$M_\odot$ star with $\Delta G/G_{\rm N}=0$ (black solid), $\Delta G/G_{\rm N}=0.05$ (red dashed), and $\Delta G/G_{\rm N}=0.1$ (blue dotted). The black dashed lines bound the instability strip.  }
    \label{fig:HR}
\end{figure}

\begin{equation}\label{eq:B}
\Delta \log(L) \simeq B \log(1+\Delta G/G_{\rm N}).
\end{equation}
The value of the slope $B$ depends on the mass of the Cepheid and whether it is observed at the second or third crossing of the instability strip,\footnote{We define the second/third crossing as the second/third time a star crosses the instability strip. The first crossing refers to the brief period where the star has exited the main sequence but not yet become a red giant. This phase is shorter than the other two and not typically observed.} as quantified in Table~\ref{tab:B} (reproduced from \cite{rho_DM}). A schematic depiction of how unscreening each region of the Cepheid affects the PLR is shown in Fig.~\ref{fig:stairway_to_heaven} (right).

\begin{table}
  \begin{center}
  \small\addtolength{\tabcolsep}{-5pt}
    \begin{tabular}{|c|c|c|}
      \hline
       $M_\text{ceph}/M_\odot$ & Slope at $2^\text{nd}$ crossing & Slope at $3^\text{rd}$ crossing\\ 
      \hline
    \rule{0pt}{3.5ex}
      7 & 4.45 & 3.79\\
    \rule{0pt}{3.5ex}
      8 & 4.34 & 3.58\\
    \rule{0pt}{3.5ex}
      9 & 4.18 & 3.46\\
      \rule{0pt}{3.5ex}
      10 & 4.00 & 3.48\\
      \rule{0pt}{3.5ex}
      11 & 3.81 & 3.58\\
      \rule{0pt}{3.5ex}
      12 & 3.67 & 3.92\\
      \rule{0pt}{3.5ex}
      13 & 3.58 & 3.95\\
      \hline
    \end{tabular}
  \caption{Slope $B$ of the $\Delta \log(L) - \log(1+\Delta G/G_{\rm N})$ relation for a range of Cepheid masses measured at the second or third crossing of the instability strip.}
  \label{tab:B}
  \end{center}
\end{table}

\section{Methodology}
\label{sec:method}

In this section we construct a model for $H_0$ including a partially-screened fifth force. Rather than repeat the full inference of $H_0$ with additional modified gravity parameters---which would require non-public data, an extensive exploration of subtleties involving the compatibility of the datasets and treatment of outliers, and a precise prior formulation of the screening models under investigation---we opt simply to modify the R19 result according to the screening properties of the Cepheids in various scenarios. This ensures that our results agree with R19 in the fiducial case of $\Lambda$CDM (or equivalently all galaxies fully screened), and we caution that therefore all the assumptions present in that analysis are included here. Sec.~\ref{sec:models} describes our methods for calculating the screening proxies, while Sec.~\ref{sec:H0_overall} models the $H_0$ posterior in a given scenario with all uncertainties folded in. This approach is model-independent and applies in full generality to any scenario in which Cepheids in the distance ladder samples experience different effective strengths of gravity. It could also be applied to scenarios in which periods and luminosities vary for other reasons. In Appendix \ref{Appendix:SN} we consider the additional possibility of the fifth force affecting SNe, which does not however affect our conclusions. 

The reader uninterested in the details of the method may safely skip now to Sec.~\ref{sec:results}, which presents our results.

\subsection{Calculating fifth force strength: The screening models}
\label{sec:models}
%

In Sec.~\ref{sec:heuristic} we laid out the screening proxies that we consider for segregating Cepheids into screened and unscreened subpopulations. Proxies \eqref{it:1}--\eqref{it:3} are determined using the screening maps of \cite{Desmond}, which synthesises galaxy survey data, halo catalogues from N-body simulations and analytic and numerical structure formation models (based partly on the BORG algorithm~\cite{Jasche, Lavaux}) to reconstruct the gravitational field out to a distance $\sim$250 Mpc. Full details can be found in the original paper. For illustration, we will consider three values for $\lambda_C$ or $R_\text{max}$---$0.5, 5$, and $50$ Mpc---which we indicate with a subscript to $\Phi$, $a$, or $K$. Proxies \eqref{it:4} and \eqref{it:5}, along with their uncertainties, are extracted from the NASA Extragalactic Database (NED) and Extragalactic Distance Database (EDD) respectively ($L_\text{gal}$ by means of the absolute visual magnitude $M_V$). We use pre-digital $W_{20}$ values for homogeneity, as many galaxies do not have digital values recorded.\footnote{A few R16 galaxies are missing information in these databases. We replace the $V$-band magnitude of NGC 4536 with the NIR magnitude, and the predigital $W_{20}$ for NGC 4038 and NGC 5917 by HIPASS measurements, also from EDD, which are similarly normalised. Galaxies for which magnitude uncertainties are not listed in NED are assigned $\Delta M_V = 0.2$.} For the MW, we estimate $M_V$ and hence $L_\text{gal}$ through a linear correlation of $M_V$ and stellar mass $M_*$ (as recorded for 145,155 galaxies in the NASA Sloan Atlas\footnote{\url{http://nsatlas.org/}}) assuming $M_*^\text{MW}=6.08 \times 10^{10} M_\odot$ \cite{MW_mass}. This yields $M_V^\text{MW}=-20.6$, which roughly agrees with the result of \cite{Karachentsev}.\footnote{Several alternative methods exist for determining the virial properties of the MW, and alternative calibrations of its magnitude and linewidth are possible. Our values are however roughly consistent with literature results (e.g. \cite{MW_1, MW_2, MW_3, MW_4, Karachentsev}).} We estimate $W_{20}$ for the MW through the $V_\text{flat}$--$W_{20}$ relation of \cite{Lelli} assuming $V_\text{flat}^\text{MW} = 220$ km/s, and conservatively assign $\Delta M_V = 0.5$, $\Delta W_{20} = 50$ km/s. Proxy \eqref{it:6} is determined from \eqref{it:4} using the technique of halo abundance matching (AM)~\cite{Kravtsov,Conroy,Behroozi_2010,Moster}, in particular the best-fit model of \cite{Lehmann}. This probabilistically assigns a galaxy of given $M_V$ to a halo from an N-body simulation (in our case \textsc{DarkSky-400} \cite{DarkSky} post-processed with \textsc{Rockstar} \cite{rockstar}) by matching the magnitude function to the function of a halo proxy formed from a combination of halo virial mass and concentration designed to maximise agreement with clustering statistics.\footnote{By using the $\Lambda$CDM halo mass function we assume our fifth force models to have negligible effect on cosmic structure formation and the density profiles of halos. For the small subset of our screening proxies for which this has been directly tested with modified gravity N-body codes, fifth force strengths and ranges that do lead to significant changes to these observables have already been ruled out \cite{Shi, Fontanot, Lombriser_rev}. Further, the $\Delta G/G_{\rm N}$ values that we find to be necessary for reconciling the R19 and \textit{Planck} results, $\mathcal{O}(0.1)$, are lower than those probed cosmologically.} Finally, proxy \eqref{it:7} is calculated by modelling each host galaxy's halo by an Navarro-Frenk-White (NFW) profile---with mass and concentration derived from AM, as above---and then sampling this density distribution at the position of a Cepheid within the galaxy. The locations of the Cepheids are taken from R16 Table 4. Note however that the full 3D distance $r$ of a Cepheid from the centre of its halo is not known, but only the projected 2D distance $R$ from the centre of the galaxy in the plane of the sky. We assume $r=\sqrt{3/2}\:R$, corresponding on average to random orientation of the line of sight with respect to the displacement of the Cepheid from the halo centre, and take the halo centre to be at the quoted galaxy position.

Given the three variants (values of $R_\text{max}$) that we consider for proxies \eqref{it:1}--\eqref{it:3} this gives 13 proxies in total. Although clearly correlated by common sources in the mass distribution both of the halos themselves and the surrounding environment, these are intended to largely span the gravitational parameter space of the galaxies. Other proxies may be formed by combination: for example, in thin-shell screening the strength of the fifth force is determined by the total Newtonian potential \cite{Cabre2012, Zhao_1, Zhao_2}, which is the sum of \eqref{it:1} and a simple function of \eqref{it:5} or \eqref{it:6}. This makes it easy to extend or approximately interpolate our results for other screening models.

Our determination of each of these quantities is inherently probabilistic, and a potentially important source of uncertainty in our estimate of $H_0$ stems from uncertainties in the galaxies' screening properties under a given model. We therefore make sure to propagate fully the uncertainties in the inputs into probability distributions for the final proxy values. The pipeline of \cite{Desmond} naturally produces probability distributions for $\Phi$, $a$ and $K$ by Monte Carlo sampling those of the inputs to that model. Uncertainties in $M_\text{vir}$ and concentration are derived by first generating 200 AM mock catalogues (sampling the uncertainty in the galaxy-halo connection induced by the AM scatter parameter \cite{Lehmann}), and then for each model realisation drawing a random magnitude for the galaxy from a normal distribution of mean $M_V$ and standard deviation $\Delta M_V$ and assigning to it the halo from a randomly-chosen AM mock that is nearest to it in magnitude. The uncertainties in $M_\text{vir}$ and concentration are propagated directly into $\rho_\text{DM}$ by an analogous Monte Carlo method. {This procedure includes the uncertainties on the proxy values of the MW and N4258 which set the screening threshold (see below), so that all errorbars are folded into the uncertainties on each galaxy's modified distance.}

Finally these proxies must be mapped to an effective $G$ value for each Cepheid. At the phenomenological level at which we are working this mapping is fairly free, although in keeping with the principle of screening it must give $\Delta G=0$ for proxy values indicating a sufficiently dense environment. Specific theories impose particular requirements on the threshold value of the proxy and the value of $\Delta G$ in the unscreened regime: e.g. in Hu-Sawicki $f(R)$ \cite{Hu07}, a common benchmark model for chameleon screening, an object is unscreened if $\log(|\Phi|/c^2) < 3/2 \: f_{R0}$ ($f_{R0}$ denotes the background value of the scalar field), in which case $\Delta G/G_{\rm N}=1/3$. In this work we consider the optimistic case in which any object with a proxy value lower than that of both N4258 and the MW is unscreened:
\begin{equation}\label{eq:DGDEF}
  G=
    \begin{cases}
      \left(1+\frac{\Delta G}{\GN}\right)\GN & p<p_\text{crit}\\
      G_\text{N} & p\ge p_\text{crit}
    \end{cases}   
\end{equation}
for any proxy $p$, with $p_\text{crit} = \text{min}(p_\text{MW}, p_\text{N4258})$. As the MW is required to be screened in any case (both to satisfy independent local tests of gravity and to ensure that the PLR calibrated by Cepheids with parallax is unchanged), this gives the largest possible unscreened fractions and hence the greatest possible reduction in $H_0$. In e.g. chameleon screening, the threshold value would not be related to $p_\text{MW}$ or $p_\text{N4258}$ but rather the self-screening parameter of the theory, $\chi$ (e.g.~\cite{Burrage:2016bwy}). For possibly viable values of this, $\lesssim 10^{-7}$, all the galaxies in our sample would self-screen (see also Sec.~\ref{sec:disc}). Thus while our assumptions suffice to give a proof of principle of the effectiveness of our mechanism at altering $H_0$ constraints, further modelling may be required for a fully satisfactory solution within the context of a given theory. For the case of $\rho_\text{DM}$ we set $p_\text{crit} = 10^7 \: M_\odot$ kpc$^{-3}$, the local dark matter density \cite{Read, Salas}.\footnote{An alternative estimate of $\rho_\text{local}$ from \textit{Gaia} suggests a slightly higher value of $1.6\times10^7 \: M_\odot$ kpc$^{-3}$ \cite{Buch}. However, we also require the bulk of the observed Cepheids in the MW to be screened so that they calibrate the Newtonian PLR. The most distant such Cepheids are $\sim$1.5$\times$ further from the galactic centre than the Sun \cite{R18}, so assuming $\rho_\text{DM} \propto r^{-1}$ within this region suggests that we need $\rho_\text{DM,crit} \lesssim 1\times10^7 \: M_\odot$ kpc$^{-3}$ to screen them all.}

For simplicity we assume that all unscreened Cepheids have a fixed value of $\Delta G/G_\text{N}$, independently of $p$. Note that in general $\Delta G/G_\text{N}$ may be an increasing function of the difference between a galaxy's proxy value and that of the MW or N4258, and may be expected to transition smoothly between screened and unscreened regimes rather than discretely. As the proxy values for the MW and N4258 are determined simultaneously with those of the other galaxies, these assumptions enable us to determine $\Delta G$ for each Cepheid for each Monte Carlo model realisation. We now describe how to propagate this information into $H_0$.

\subsection{$H_0$ model and sampling methods}
\label{sec:H0_overall}

\subsubsection{Cepheid distances}
The first step is to determine new constraints on the distances to the extragalactic Cepheid hosts. In the absence of information to the contrary we assume that in the $\Lambda$CDM case each Cepheid $j$ in galaxy $i$ implies the same distance to the galaxy, $d_{ij}$, so that the overall inferred distance $D_i = d_{ij}$. Under a fifth force model the Cepheid distances are altered to $\bar{d}_{ij}$, as we quantify below, so that the true galaxy distances $\bar{D}_i$ are an average over the distances $\bar{d}_{ij}$ implied by each of the individual Cepheids, weighted by the inverse square of their magnitude uncertainties $\Delta m_{ij}$ (period uncertainties are subdominant and not quoted in R16). This ensures that galaxies with smaller measurement uncertainties contribute more to $\bar{D}_i$ than those whose properties are less well known.

Combining Eqs.~\eqref{eq:LchangeP} and~\eqref{eq:B},
\begin{align}
\log(\bar{L}_{ij}) &= \log(L_{ij}) + \frac{A}{2} \log\left(1+\frac{\Delta G_{ij}}{G_{\rm N}}\right) + B \log\left(1+\frac{\Delta G_{ij}}{G_{\rm N}}\right), \nonumber \\
\bar{L}_{ij} &= \left(1+\frac{\Delta G_{ij}}{G_{\rm N}}\right)^{\frac{A}{2} + B} \: L_{ij},
\end{align}
where $\Delta G_{ij}$ is the enhancement to Newton's constant experienced by the Cepheid, $\bar{L}_{ij}$ is the Cepheid's true luminosity and $L_{ij}$ is that inferred from a Newtonian analysis. The distance to the Cepheid is determined by $d \propto (L/f)^{1/2}$, where $f$ the measured flux, and is therefore increased according to
\begin{equation}\label{eq:d}
\bar{d}_{ij} = \left(1+\frac{\Delta G_{ij}}{G_{\rm N}}\right)^\frac{A+2B}{4} d_{ij} \equiv k_{ij} \: d_{ij},
\end{equation}
where we define $k_{ij}$ as the total factor by which the distance is changed. To account for the possible variation in the $\log(\bar{L})$--$\log(1+\Delta G/G_{\rm N})$ relation due to varying $B$ (Sec.~\ref{sec:F5_cepheids}), for each Monte Carlo realisation of our model we separately assign each Cepheid a $50\%$ chance of being observed at second or third crossing and an equal probability of having an integer mass in the range $7$--$13M_\odot$, and adopt the corresponding value from Table~\ref{tab:B}. This effectively marginalises over our ignorance of these properties and propagates the corresponding uncertainties into the distance estimates. As can be seen from Eq.~\eqref{eq:d} and Fig.~\ref{fig:stairway_to_heaven} (right), $B>0$ shifts the unscreened PLR in the same direction as the reduction in period ($A>0$) and hence amplifies the overall bias in the inferred distances. We remind the reader that the $G$-dependence of the Cepheid magnitude is likely \emph{not} to be present in chameleon and similar models that screen the Solar System, since $G$ rapidly reverts to $G_{\rm N}$ inside a Cepheid's core. This corresponds to $B=0$.

Combining the contributions from all Cepheids in a galaxy, weighted by their uncertainties, the distance to a given host is modified according to
\begin{equation}\label{eq:d_avg}
\bar{D}_i = \left(\frac{\sum_j k_{ij} \Delta m_{ij}^{-2}}{\sum_j \Delta m_{ij}^{-2}}\right) D_i \equiv K_i D_i.
\end{equation}
The fractional scatter among the distance estimates of the individual Cepheids is given by
\begin{equation}\label{eq:d_err}
\sigma (d_i) = \left(\frac{\sum_j k^2_{ij} \Delta m_{ij}^{-2}}{\sum_j \Delta m_{ij}^{-2}} - K_i^2\right)^{1/2}.
\end{equation}
We estimate the uncertainty on $\bar{D}_i$ by summing $\sigma (d_i)$ in quadrature with the fractional uncertainty on $D_i$:
\begin{equation}\label{eq:d_var}
\Delta \bar{D}_i = \bar{D}_i \left((\Delta D_i/D_i)^2 + \sigma (d_i)\right)^{1/2}.
\end{equation}
$D_i$ and $\Delta D_i$ are calculated from the independent, Cepheid-only distance moduli listed in R16 table 5.\footnote{Note that this always yields a larger fractional uncertainty on $\bar{D}_i$ than $D_i$, since the assumption that all the Cepheids in the galaxy yield the same distance in $\Lambda$CDM requires that the dispersion can only increase when the distances are modified. In reality the dispersion may instead be reduced. $\sigma (d_i)$ is however only nonzero when Cepheids within the same galaxy can have different $G$, which is only the case for the baryon--dark matter interaction model with screening governed by $\rho_\text{DM}$. Even then it is typically small relative to $\Delta D_i/D_i$, and our conclusions are unaffected by removing the $\sigma (d_i)$ and fixing the fractional distance uncertainties.}

\subsubsection{Supernova calibration}

To propagate the distance estimates into a constraint on $H_0$, the absolute SN magnitudes are calibrated for galaxies with both SNe and Cepheid distances, and then the SN magnitude--redshift relation is extended to greater distance with a cosmological SN sample. We adopt a similar philosophy for treating this part of the distance ladder. The Newtonian ($\Delta G=0$) distances $D_i$ imply the SN absolute magnitudes $M_\text{SN}^{\rm R16}$ quoted in R16, which, combined with the improved measurement of MW Cepheid parallaxes from \textit{Gaia} in R18 and additional LMC Cepheids in R19, imply $H_0 = H_0^\text{R19} \equiv 74.03 \pm 1.42$ km/s/Mpc. $H_0$ is calculated according to (R11, R16)
\begin{equation}\label{eq:H0_1}
H_0 = 10^{M_\text{SN}/5 + 5 + a_B} \; \text{km s}^{-1} \text{Mpc}^{-1},
\end{equation}
where $a_B$ is the normalization of the magnitude--redshift relation of the cosmological SN sample, for which we retain the R16 value $a_B = 0.713 \pm 0.002$. Writing this in terms of the apparent magnitude and galaxy distance, the true $H_0$ value implied by galaxy $i$ is
\begin{align}\label{eq:H0_2}
\bar{H}_{0,i} &= 10^{m_\text{SN,i}/5 + 6 + a_B} \: \frac{\text{pc}}{\bar{D}_i} 
\: \text{km s}^{-1} \text{Mpc}^{-1} = \frac{H_0^\text{R19}}{K_i} \: .
\end{align}

\subsubsection{Overall inferred $H_0$}
Combining the modification in Eq.~\eqref{eq:H0_2} over all the galaxies in the sample yields a new best-fit estimate for $H_0$
\begin{equation}\label{eq:H0_4}
\bar{H}_0 = H_0^\text{R19} \frac{\sum_i K_i^{-1} {\Delta \bar{H}_{0,i}^{-2}}}{\sum_i {\Delta \bar{H}_{0,i}^{-2}}},
\end{equation}
where $\Delta H_{0,i}$ is the fractional uncertainty in the $H_0$ estimate of galaxy $i$ derived by propagating uncertainties in Eq.~\eqref{eq:H0_2}. This differs between the R19 inference and ours according to the best-fit distances to the galaxies and their uncertainties:
\begin{equation}\label{eq:delta_H0}
\begin{split}
&\Delta H_{0,i}^2 = \ln(10)^2 \: (\Delta m_\text{SN,i}^2/25 + \Delta a_B^2) + (\Delta D_i/D_i)^2, \\
&\Delta \bar{H}_{0,i}^2 = \ln(10)^2 \: (\Delta m_\text{SN,i}^2/25 + \Delta a_B^2) + (\Delta \bar{D}_i/\bar{D}_i)^2.
\end{split}
\end{equation}
We estimate the uncertainty in the overall $\bar{H}_0$ as
\begin{equation}\label{eq:dH0}
\Delta \bar{H}_0 = \bar{H}_0 \left[\left(\frac{\Delta H_0^\text{R19}}{H_0^\text{R19}}\right)^2 + \left(\frac{1}{\sum_i {\Delta \bar{H}_{0,i}^{-2}}}  - \frac{1}{\sum_i {{\Delta H_{0,i}^{-2}}}}\right)\right]^{1/2}.
\end{equation}
This formula modifies the fractional uncertainty on $H_0^\text{R19}$ according to the variance in the estimated $H_0$ values among the galaxies, derived from propagating errors in Eq.~\eqref{eq:H0_4}. Note that since $\Delta \bar{D}_i > \Delta D_i$, the second term within the square root is always positive and hence the fractional error on $H_0$ is increased by the addition of partial screening.  We emphasise that our easing of the $H_0$ tension is nevertheless driven by changes to the mean, rather than to the uncertainty.

Finally, we note that there may be uncertainty in the $\Delta G_{ij}$ values themselves due to uncertainty in the proxies that determine whether or not an object is screened. We propagate this into our $H_0$ constraint by performing 10,000 random Monte Carlo draws from the probability distributions describing the proxy for a given model (Sec.~\ref{sec:models}), in each case calculating $\bar{H}_0$ and $\Delta \bar{H}_0$ according to Eqs.~\eqref{eq:d}-\eqref{eq:dH0}. We estimate the true $H_0$ for this model realisation by scattering $\bar{H}_0$ by $\Delta \bar{H}_0$. The set of 10,000 $H_0$ values thereby obtained describes the $H_0$ posterior for this model, accounting fully for all sources of uncertainty. We summarise this posterior by its mode $\hat{H}_0$, standard deviation $\Delta \hat{H}_0$ and minimal (asymmetric) interval enclosing $68\%$ of the model realisations. We use the latter to calculate separate $1\sigma$ upper and lower uncertainties on $\hat{H}_0$, $\Delta \hat{H}_0^{+/-}$, to account for non-Gaussianity in the posterior. We also combine $\hat{H}_0$ and the uncertainty into an overall discrepancy with \textit{Planck} in order to quantify the ability of a given screening model to reduce the $H_0$ tension. For consistency with literature estimates of the discrepancy in other models, and for simplicity, we calculate this using the symmetrised errorbar $\Delta \hat{H}_0$:
\begin{equation}\label{eq:sig_H0}
\sigma_{H_0} \equiv \frac{\hat{H}_0 - H_\text{CMB}}{\sqrt{\Delta \hat{H}_0^2 + \Delta H_\text{CMB}^2}},
\end{equation}
where $H_\text{CMB} = 67.4$ km s$^{-1}$ Mpc$^{-1}$ and $\Delta H_\text{CMB} = 0.5$ km s$^{-1}$ Mpc$^{-1}$. This is to be compared with $\sigma_{H_0} = 4.4$ in the $\Lambda$CDM case $\Delta G_{ij} = 0$.


\section{Results}
\label{sec:results}

In this section we show the screening properties of the R16 Cepheids and their hosts, describe consistency tests within the distance ladder data that limit the possible strength of a fifth force, and quantify the change in best-fit $H_0$ within our formalism.

\subsection{Screening properties of Cepheids and their hosts}
\label{sec:screening_props}

\begin{figure*}
  \centering
  \includegraphics[width=0.49\textwidth]{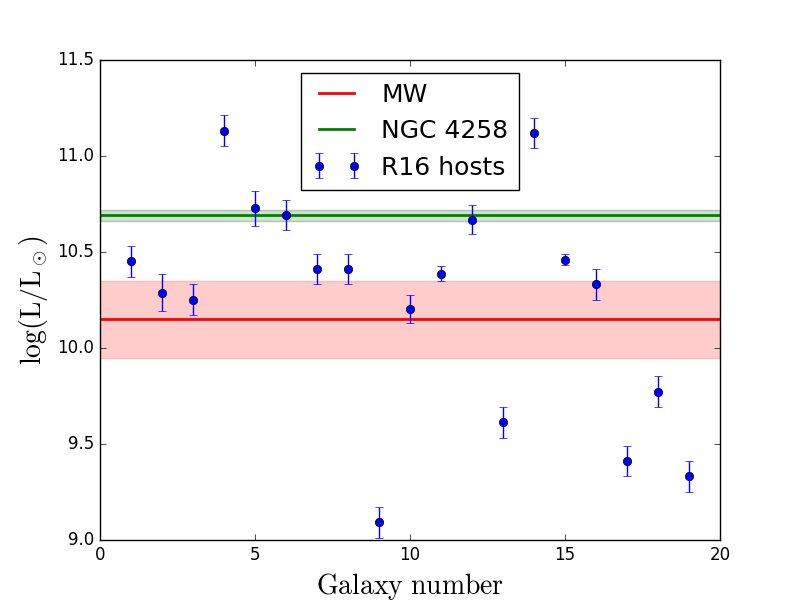}
  \includegraphics[width=0.49\textwidth]{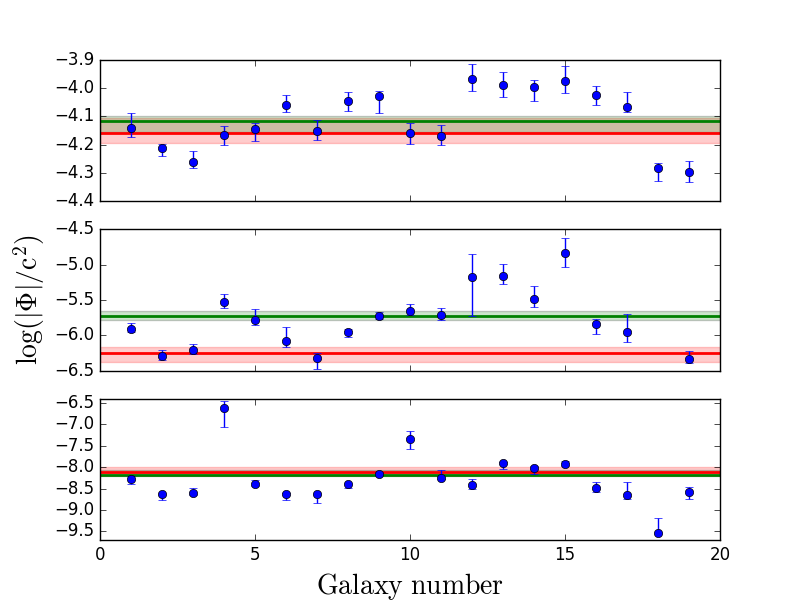}
  \includegraphics[width=0.49\textwidth]{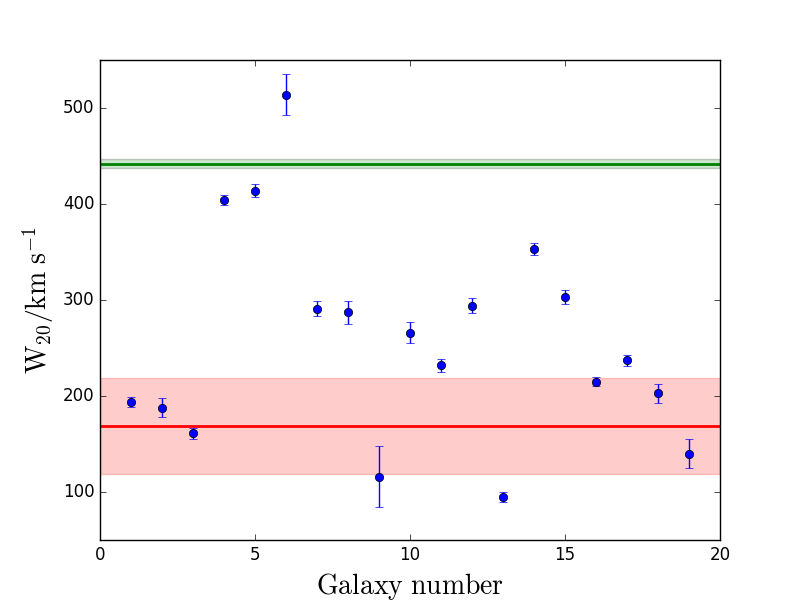}
  \includegraphics[width=0.49\textwidth]{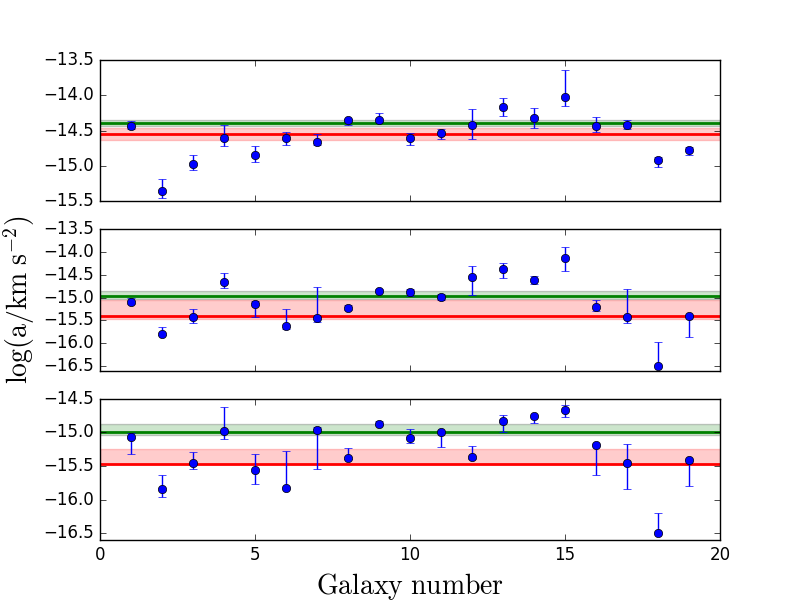}
  \includegraphics[width=0.49\textwidth]{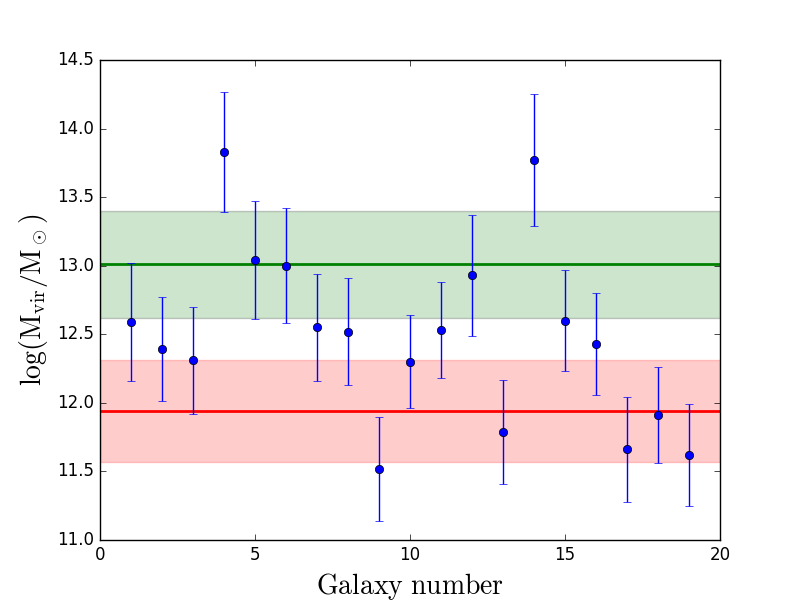}
  \includegraphics[width=0.49\textwidth]{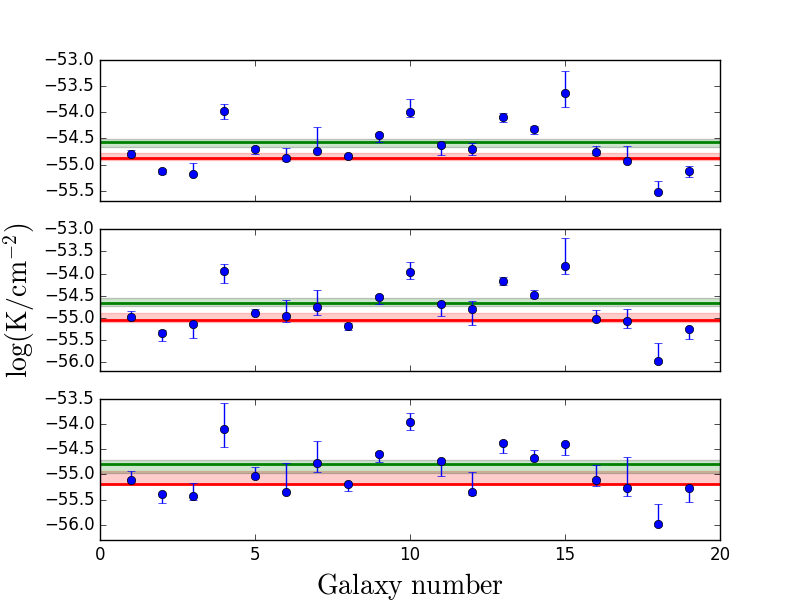}
  \caption{The screening proxies of luminosity, mass estimated through hydrogen gas, halo virial mass and external potential, acceleration and curvature over the Cepheid host galaxies in the R16 sample. The anchors, MW and N4258, are shown separately by the red and green horizontal lines. Galaxies lying below these lines have weaker internal gravitational fields than the MW and N4258, and may therefore be unscreened. The ordering of the galaxies is as in Table~\ref{tab:galprops1}, starting with M101. The subpanels of the environmental screening plots (right column) correspond to the three different distances $R_\text{max}$ out to which we include contributions from masses: $0.5$ Mpc (lower), $5$ Mpc (middle) and $50$ Mpc (upper). The asymmetric errorbars indicate the minimal width enclosing $68\%$ of the Monte Carlo model realisations.}
  \label{fig:proxies}
\end{figure*}

In Fig.~\ref{fig:proxies} we show the proxy values over all the galaxies as determined by the procedure of Sec.~\ref{sec:method}. The errorbars show the minimal width enclosing $68\%$ of the model realisations, the red line with shaded errorbar shows the value for the MW and the green line the value for N4258. Only galaxies below both the red and green lines have the potential for being unscreened if one requires that both the MW and N4258 Cepheids calibrate the Newtonian or screened PLR. We see that the MW is in almost all cases harder to screen than N4258, and hence it is this galaxy that effectively sets the threshold. The properties themselves are listed in tables in Appendix~\ref{sec:app1}.

Fig.~\ref{fig:densities} shows the distribution of dark matter densities $\rho_\text{DM}$ at the positions of the Cepheids in each of the extragalactic hosts separately. The red line gives the dark matter density at the position of the Sun \cite{Read}, which we take to be the screening threshold in the $\rho_\text{DM}$ model: all Cepheids to the left of this line are therefore considered unscreened.

In the first column of Table~\ref{tab:H0} (Appendix ~\ref{sec:app1}) we show the unscreened fraction of Cepheids corresponding to each of the proxies.\footnote{Note that in models where Cepheids do not screen differentially within a galaxy (all proxies except $\rho_\text{DM}$), the galaxies are simply weighted by the number of Cepheids they contain in calculating this fraction.} This shows the MW to be typical in the full set of Cepheid hosts but somewhat less dense (or in a less dense environment) than the average, so that when the MW is screened typical unscreened fractions are $\sim$30\%. The largest unscreened fraction is obtained for $\Phi_{0.5}$, indicating that the immediate ($<0.5$ Mpc) environment of the MW is more dense than that of the majority of the other hosts, and vice versa for models based on the luminosity or mass estimated through hydrogen gas ($L_\text{gal}$ and $W_{20}$). A relatively high unscreened fraction is also achieved when segregating the Cepheids by the local dark matter density. As expected, the larger the unscreened fraction $f$ the greater the effect on $H_0$ for given $\dG$. That there is little scatter in the $\sigma_{H_0}(f)$ relation indicates that the total number of unscreened Cepheids is the most important quantity for setting the $H_0$ bias, while which those Cepheids are is secondary. Thus the precise mechanism of screening is not particularly important; what matters is mainly the fraction of Cepheids in extragalactic hosts that feel the fifth force.

\begin{figure}
  \centering
  \includegraphics[width=0.5\textwidth]{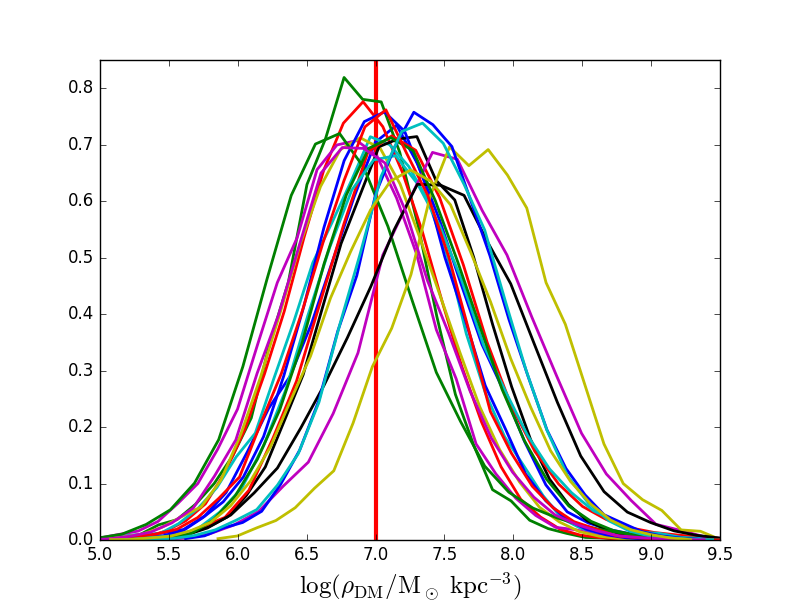}
  \caption{Normalised frequency distributions of dark matter densities $\rho_\text{DM}$ at the positions of the R16 Cepheids within their hosts, estimated using the halo properties from Table~\ref{tab:galprops2}. Each curve corresponds to a different galaxy. The local dark matter density, $10^7 M_\text{sun}$ kpc$^{-3}$, is shown by the vertical red line: Cepheids to the left of this line live in lower density regions than the Solar System and may therefore be unscreened in the baryon--dark matter interaction model.}
  \label{fig:densities}
\end{figure}

\subsection{The viable range of $\Delta G/G_{\rm N}$: Consistency tests within the distance ladder}
\label{sec:tests}

Before showing results for the impact of our fifth force models on $H_0$, we describe three ways in which they may be constrained by distance ladder data. This will indicate the maximum viable values of $\dG$, which we find to be $\sim$0.15 for models that only unscreen Cepheid envelopes, and $\sim$0.04 for models that unscreen Cepheid cores as well.

\begin{figure*}
  \centering
  \includegraphics[width=0.49\textwidth]{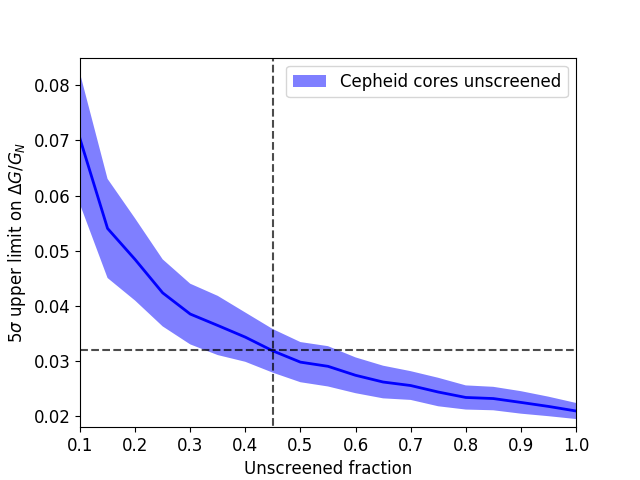}
  \includegraphics[width=0.49\textwidth]{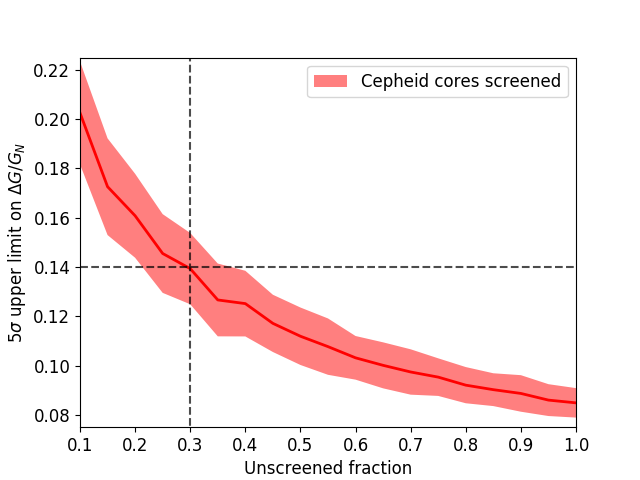}
  \caption{Upper limit on $\Delta G/G_\text{N}$ from the comparison of Cepheid and TRGB distance estimates to 51 galaxies from NED-D, as a function of the fraction of unscreened galaxies. We show separately the cases of Cepheids entirely unscreened (\emph{left}) and only Cepheids' envelopes unscreened (\emph{right}). Dashed lines indicate typical unscreened fractions as shown in Table~\ref{tab:H0}.}
  \label{fig:trgb}
\end{figure*}

Our principal test compares distance estimates to the same galaxy by means of the Cepheid PLR vs tip-of-the-red-giant-branch (TRGB) luminosity. As we have seen, unscreening Cepheids results in the standard analysis underestimating the distances to their hosts. Further, as shown by means of numerical simulation in \cite{rho_DM}, at $\dG \gtrsim 0.1$ the hydrogen-burning shell that sources the TRGB radiation becomes unscreened, reducing its luminosity. This causes their distance to be overestimated. Combining eqs. 40 and 41 of \cite{rho_DM}, we find that the TRGB distance modification for general degree of screening is well-fit by
\begin{equation}\label{eq:d_trgb}
\frac{D_\text{true}}{D_\text{GR}} = 1.021 \left(1-0.04663 \left(1+\frac{\Delta G}{G_\text{N}}\right)^{8.389} \right)^{1/2},
\end{equation}
where $D_\text{GR}$ is the conventional result assuming General Relativity (GR). The Cepheid and TRGB modifications act in opposite directions, which means that in the presence of a screened fifth force the two distance estimates would be expected to systematically disagree, with TRGB normalised higher. We calculate the constraints this imposes on $\Delta G/G_\text{N}$ by updating the analysis of \cite{Jain} with the latest data from the Nasa Extragalactic Distance Database (NED-D)\footnote{\url{https://ned.ipac.caltech.edu/Library/Distances/}.} \cite{NED-D}, and generalising from the specific case of chameleons considered in that work. We find 51 galaxies out to 35 Mpc with both Cepheid and TRGB distance estimates. For each one, we calculate the uncertainty-weighted average Cepheid and TRGB distance, and hence the fractional difference $(D_\text{ceph,obs}-D_\text{trgb,obs})/D_\text{trgb,obs} \equiv \Delta D/D_\text{obs}$ and corresponding uncertainty $\sigma_{\Delta D/D}$. This is to be compared to the expectation from Eqs.~\eqref{eq:d} and~\eqref{eq:d_trgb}:
\begin{align}\label{eq:dd}
\left\langle \frac{\Delta D}{D} \right\rangle = &\left(1+\frac{\Delta G}{G_\text{N}}\right)^{-\frac{A+2B}{4}} 1.021 \left(1-0.04663 \left(1+\frac{\Delta G}{G_\text{N}}\right)^{8.389} \right)^{1/2} \\ \nonumber
&-1,
\end{align}
where $A$ and $B$ again encode the contributions to the distance modification from unscreening the Cepheid envelope and core respectively, and angle brackets denote theoretical expectation at given $\Delta G/G_\text{N}$. We perform the inference with a standard Gaussian likelihood
\begin{equation}
\ln L\left(\frac{\Delta D}{D}_\text{obs} \: \bigg\rvert \: \frac{\Delta G}{G_\text{N}}, \sigma_\text{n}\right) = -\frac{(\frac{\Delta D}{D}_\text{obs} - \langle \frac{\Delta D}{D} \rangle)^2}{2 \sigma_\text{tot}^2} - \frac{1}{2} \ln(2 \pi \sigma_\text{tot}^2),
\end{equation}
where $\sigma_\text{tot}^2 \equiv \sigma_{\Delta D/D}^2 + \sigma_\text{n}^2$, and $\sigma_\text{n}$ is an additional noise term accounting for astrophysical contributions to the variance in $D_\text{ceph/trgb}$ not captured by the measurement uncertainty. Eq.~\eqref{eq:dd} only gives the expectation for unscreened galaxies however; in screened galaxies we instead have $\langle \Delta D/D\rangle = 0$. To keep the test completely general as regards screening mechanism, we assume that a fraction $f$ of the 51 galaxies in the sample are unscreened. We randomise which galaxies these are and repeat the inference of $\{\Delta G/G_\text{N}, \: \sigma_\text{n}\}$ 50 times for each $f$ in the range $0.05-1$, in order to estimate the variance in the result as different galaxies in the sample are unscreened. Marginalising over $\sigma_\text{n}$, which is well-constrained by the variance in $\Delta D/D_\text{obs}$ among galaxies, we find all $\Delta G/G_\text{N}$ posteriors to be consistent with 0. We calculate from these the $5\sigma$ upper limit on $\Delta G/G_\text{N}$ as a function of $f$, assuming either that Cepheids are entirely unscreened ($A=1.3$, $B\simeq3.85$ [a simple average of the values in Table~\ref{tab:B}]) or that only Cepheid envelopes are unscreened ($A=1.3$, $B=0$). The results are shown in separate panels of Fig.~\ref{fig:trgb}, where the shaded region represents the $1\sigma$ uncertainty in the upper limit on $\Delta G/G_\text{N}$ due to ignorance of which galaxies are unscreened at given $f$. (Note that in principle one could remove this uncertainty by calculating the screening level of each Cepheid and TRGB under a given model.) The dotted lines indicate typical $f$ values from the smaller R16 sample (Table~\ref{tab:H0}): for the $B\ne0$ case we show the result for the $\rho_\text{DM}$ screening model ($f=0.45$), corresponding to $\Delta G/G_\text{N} \lesssim 0.03$, while for the $B=0$ case we highlight $f=0.3$, which yields $\Delta G/G_\text{N} \lesssim 0.14$.

Using the $5\sigma$ bound on $\dG$ from this test may seem excessively lenient. However, the test is unlikely to be statistically limited because it suffers from a number of potentially significant systematic uncertainties that are difficult to assess. The measurements in NED-D incorporate a diverse range of separate studies with different assumptions for the photometric pipeline, the metallicities of the Cepheids, the intrinsic and external extinction and reddening and the zero-points for calibrating the Cepheid and TRGB distances. In addition, multiple studies may reduce the same data in different ways, so that the listed measurements are not independent. It is beyond the scope of our work to assess the robustness of this test in detail or attempt to mitigate the systematics; we simply note that even the $5\sigma$ limiting values could be fully allowed were all errors taken into account. Further discussion of these and related issues may be found in \cite{Beaton_1, Beaton_2, Beaton_3, Beaton_4, F19}.

A second test compares the brightnesses of SNe in different galaxies. If SNe are standardizable candles they should all have the same universal absolute magnitude after rescaling the width of the light curve. Separating the galaxies into screened and unscreened subsamples under a given screening model, it should be possible to reconstruct this for each subsample separately by using the appropriate Cepheid distance estimates. The closer the $\Delta G_{ij}$ to their true values, the smaller one would expect the difference between the absolute magnitudes of the screened and unscreened SNe to be.
We find a clear segregation of the $M_\text{SN}$ values of screened and unscreened galaxies when $\Delta G/G_\text{N} \gtrsim 1$ for the models in which only Cepheid envelopes are unscreened, and $\Delta G/G_\text{N} \gtrsim 0.3$ when Cepheids are entirely unscreened: this test is therefore considerably weaker than that using TRGBs described above.\footnote{Note however that the constraints from this test are not sensitive to the unscreened fraction of galaxies in a given model. Hence they may be stronger than those from the TRGB test for proxies producing very low unscreened fractions, e.g. $\Phi_5$.} We have however assumed that the SN luminosity should be the same between screened and unscreened galaxies, i.e. no fifth force effects on the SNe themselves. This is discussed further in Appendix~\ref{Appendix:SN}.

The final test examines the scatter in the PLR. When all Cepheids in a galaxy are unscreened only the normalization of the PLR changes, which is fully degenerate with a change in distance. However, galaxies containing Cepheids with a greater range of degrees of screening, and hence variance in $G_{ij}$, would be expected to have a larger scatter in their individual period--magnitude relations, as some scatter would derive from misestimation of their individual distances causing a shift in apparent magnitude. ``Correcting'' this according to the true screening model should then reduce the scatter. We do not find a significant correlation between this scatter and the fraction of unscreened Cepheids across the R16 galaxies in the $\rho_\text{DM}$ model (the only model we consider in which Cepheids within a single galaxy may be differently screened), nor do we find individual Cepheids' magnitude residuals to correlate with their degree of screening. Instead, the average PLR scatter is significantly \emph{increased} relative to the magnitude uncertainties (due to a bimodality induced between screened and unscreened Cepheids within a given galaxy) when $\dG \gtrsim 2$ for only Cepheid envelopes unscreened, or $\dG \gtrsim 0.5$ for Cepheids entirely unscreened. Taking these as approximate limits, this test is weaker than both those above.

Each of these tests may be improved with future data, and more may be developed by considering other aspects of the distance ladder. For present purposes, however, we take the maximum viable values of $\dG$ from the Cepheid--TRGB distance test.

\subsection{Effect on $H_0$}
\label{sec:H0_effect}

\begin{figure*}
  \centering
  \includegraphics[width=0.495\textwidth]{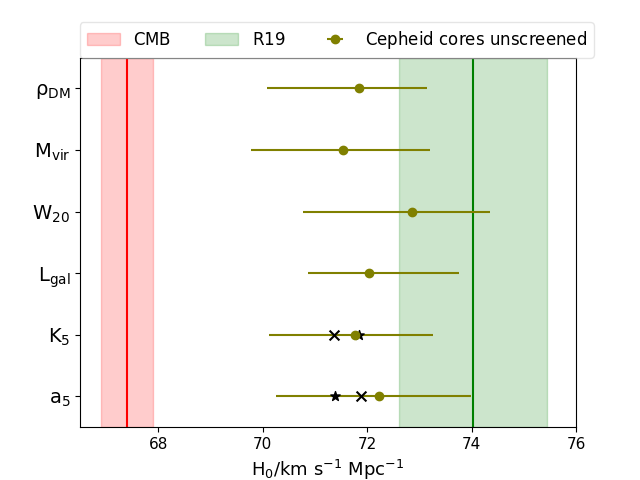}
  \includegraphics[width=0.495\textwidth]{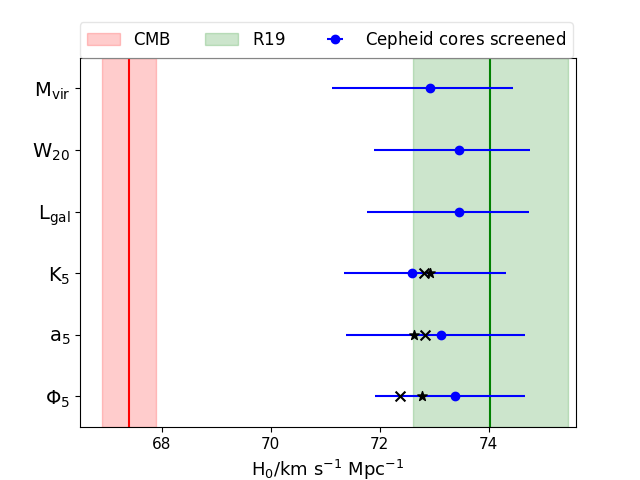}
  \caption{Reconstructed $\hat{H}_0$ values and their uncertainties for the two main screening scenarios that we consider. The points in the left panel correspond to Cepheids entirely unscreened, and in the right panel to only Cepheid envelopes unscreened. In all cases we use the $5\sigma$ maximum values for $\dG$ from the Cepheid vs TRGB distance test (Sec.~\ref{sec:tests} and Fig.~\ref{fig:trgb}), which span the range 0.03--0.06 for the left panel and 0.13--0.22 for the right (Table~\ref{tab:H0}). The red and green vertical lines and shaded regions show the best-fit $H_0$ value, and its $1\sigma$ uncertainty, from \textit{Planck} and R19 respectively. For the environmental proxies $\Phi$, $a$ and $K$ we show results for a $5$ Mpc aperture, but indicate also the result of using a $0.5$ or $50$ Mpc aperture with black crosses and stars respectively (the errorbars are similar). 
  Note that the $\rho_\text{DM}$ model must unscreen Cepheid cores so appears only in the left panel, while the converse is true for $\Phi$ assuming thin-shell screening.
  }
\label{fig:H0_proxies}
\end{figure*}

Fig.~\ref{fig:H0_proxies} shows $\hat{H}_0 \pm \Delta \hat{H}_0^{+/-}$ for Cepheid cores either unscreened ($B$ in Eq.~\eqref{eq:d} from Table~\ref{tab:B}; left) or screened ($B=0$; right). Unscreening Cepheid cores has the same effect on $\hat{H}_0$ as increasing $\Delta G/G_\text{N}$ by a factor $\sim$6$-8$. For each model we take the maximum allowed value of $\dG$ from Fig.~\ref{fig:trgb}, assuming the same unscreened fraction of galaxies in the R16 sample as the Cepheid--TRGB sample from NED-D. The $\dG$ values themselves are given in Table~\ref{tab:H0}, along with the corresponding values of $\hat{H}_0$, $\Delta \hat{H}_0$ and $\sigma_{H_0}$.

\begin{figure}
  \centering
  \includegraphics[width=0.495\textwidth]{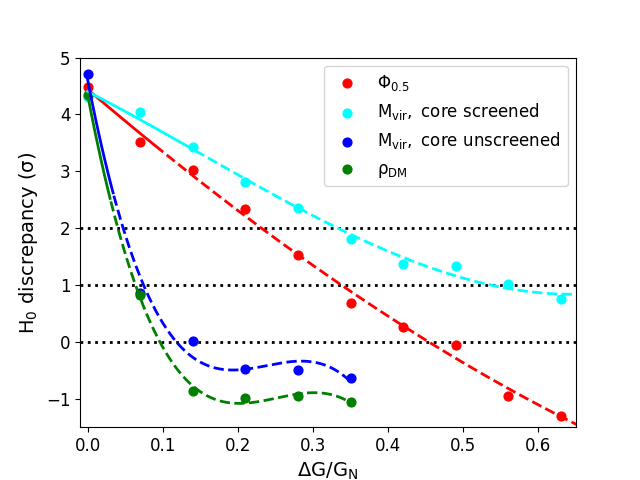}
  \caption{Distance ladder vs CMB $H_0$ discrepancy $\sigma_{H_0}$ as a function of $\Delta G/G_{\rm N}$ for four screening models: $\rho_\text{DM}$, $\Phi_{0.5}$, and $M_\text{vir}$ both with and without Cepheid luminosity modification. The lines are cubic fits to the corresponding points. We cut the $\rho_\text{DM}$ and $M_\text{vir}$ models where they are inconsistent with all the tests of Sec.~\ref{sec:tests}, and dash the lines where they enter $5\sigma$ tension with the Cepheid vs TRGB distance test.}
  \label{fig:dG_discrep}
\end{figure}

Fig.~\ref{fig:dG_discrep} shows the $H_0$ discrepancy $\sigma_{H_0}$ for four models---$\Phi_{0.5}$, $\rho_\text{DM}$ and $M_\text{vir}$ both with and without Cepheid luminosity modification---as a function of $\Delta G/G_{\rm N}$. These indicate that it is possible to remove the $H_0$ tension by means of screening given a fifth force strength $\sim$10\% that of gravity when Cepheid luminosities are altered (i.e. their cores are unscreened) and $\sim$50-100\% of gravity if only their periods are altered. These values are however in tension with the Cepheid vs TRGB distance test of Sec.~\ref{sec:tests}, and we dash the lines where this tension exceeds 5$\sigma$. Taking this constraint into account, our models with Cepheid cores unscreened can reduce the $H_0$ tension to $\sim$2$-3\sigma$, while those with Cepheid cores screened reduce it to $\sim$3$-4\sigma$.

\section{Discussion}
\label{sec:disc}

In this section we indicate caveats and possible systematics in our proposed resolution to the $H_0$ problem, and suggest ways to develop it further.

By construction, our models screen the MW and hence pass local tests of the equivalence principles, the inverse-square law, and post-Newtonian gravity \cite{Adelberger,Sakstein:2017pqi}. Their effects on cosmological observables are however less certain. Chameleon and similar models have little relation to dark energy \cite{Wang:2012kj,Lombriser:2016yzn}, while dark energy theories that screen using the kinetic or Vainshtein mechanisms require parameters such that Cepheids (and galaxies as a whole) are self-screened. Hence our implementation of these models is not cosmological but rather represents an intermediate-scale modification of GR. The baryon--dark matter interaction model \cite{Berezhiani:2016dne} and its screening proxy $\rho_\text{DM}$ have not been explored in detail. 
In~\cite{rho_DM} we derive the parameterised post-Newtonian (PPN) parameter $\gamma$ for this model, finding that it too is dark matter density-dependent. If the MW is screened, tests based on $\gamma$ would require data from less dense galaxies.

The $\dG$ parameter space for some of our proxies has already been constrained by prior analyses. For example, thin-shell models (chameleon, symmetron and dilaton) have a threshold in $\Phi$ which determines when screening is active. $\Delta G/G_\text{N}$ values as low as 0.1 for these theories have been ruled out down to $\lambda_C$ ($\simeq R_\text{max}$) $\sim 1$ Mpc 
\cite{Jain,Vikram,Desmond_letter}. 
These constraints impact the viability of screening based on external potential, internal velocity, or galaxy or halo mass.  
As we require $\Delta G/G_{\rm N} \gtrsim 0.5$ for consistency of the distance ladder and \textit{Planck} results when Cepheid cores are screened, a resolution of the $H_0$ tension is unlikely feasible within the thin-shell paradigm. Similarly, for kinetic and Vainshtein models the parameters for which stars are unscreened are already ruled out \cite{deRham:2016nuf,Sakstein:2017pqi,Sakstein:2017bws,Sakstein:2017xjx}, so that our models based on $a$ and $K$ would be required to invoke similar but not identical mechanisms.

Beyond the MW and N4258, further constraining power on the shape of the PLR derives from independent eclipsing binary distance estimates to M31 and the LMC, which we have not required to be screened. 
This is partly justified by the fact that N4258 is the most important calibration galaxy because its Cepheids have long periods and are therefore most similar to those in the SN hosts \cite{Follin_Knox}. M31 was not used as an anchor in R19 as its eclipsing binaries are early-type and therefore suffer from systematics associated with out-of-equilibrium atmospheric physics that are hard to assess. M31 is however at least as massive as the MW; it is therefore just as strongly screened and we expect that its Cepheids could be incorporated straightforwardly into our framework without loss of effectiveness at reducing $H_0$. 
There are grounds also for believing the LMC to be a less reliable anchor than N4258: it requires a larger (uncertain) metallicity correction and is somewhat more prone to photometric systematics. It is however significantly less massive than the MW, making it more likely to be unscreened at least under the models based on luminosity and internal velocity. It may also be unscreened under $\rho_{\rm DM}$, although the total density including the subhalo's own mass is of order our threshold screening value of $10^7 M_\odot$ kpc$^{-3}$ \cite{LMC_1, LMC_2}. Larger-scale environmental proxies are very similar between the LMC and MW since the former lies well within the halo of the latter. Unscreening at least some LMC Cepheids would push the calibrated PLR towards the unscreened relation, reducing the change in $H_0$. One might therefore expect a lower reconstructed $H_0$ with the LMC as sole distance anchor, which does not appear to be the case in the latest analysis (R19). 

$G > G_{\rm N}$ affects other observables besides the periods and luminosities of Cepheids, for example their light curves and spectral features. These are well measured and in good agreement with numerical predictions based on GR, and may therefore be used to further constrain the models we consider \cite{Marconi:2013tta}. We have also assumed that fifth forces do not significantly affect the other objects that help calibrate the PLR, for example the water maser in N4258 and eclipsing binaries in the LMC and M31.  In chameleon (and similar), Vainshtein and kinetic screening models, it is likely that binaries are self-screened, removing any uncertainty. One possible exception in chameleon theories is if at least one of the stars is post-main-sequence. In the case of the dark matter--baryon interaction the effects are not so clear: one would need to extend the one-body computation to the more technical two-body problem \cite{Brax:2018bow,Kuntz:2019zef,Brax:2019tcy}. If the eclipsing binaries are unscreened $G_{\rm N}$ becomes $G_{\rm N}+\Delta G$ in Kepler's third law, neglect of which results in an overestimate of the sum of the masses and hence of each individual mass. This would overestimate the luminosity, except that the stars themselves are more luminous because of their stronger self-gravity, partially compensating for this effect. A full modelling of these effects is left for future work. It may also be fruitful to compare the Cepheid results with those from other variable stars such as Miras \cite{Miras}, 
which have different internal and environmental gravitational properties and may therefore experience different degrees of screening. This could provide constraints as powerful as those of the TRGB test of Sec.~\ref{sec:tests}. The most general test may be to compare the masses of stars derived through dynamical, Hertzsprung--Russell and surface gravity methods: different dependences on $G$ would lead to systematic biases that correlate with environment.

The models could also be tested with other aspects of distance ladder data. For example, a given model predicts whether particular Cepheids are more likely to scatter above or below the PLR as a function of their screening properties. Subsets of the data may therefore be formed which differ in screening while being alike in other ways, and these would be expected to trace different PLRs and hence give systematically different results for $H_0$. This would enable further development of consistency tests along the lines of those in Sec.~\ref{sec:tests}. It would also be useful to investigate the systematic uncertainties associated with the Cepheid--TRGB distance test that we have used to bound $\dG$. This would reveal whether our $5\sigma$ limits are overly lenient so that our $H_0$ constraints are over-optimistic, or conversely whether even larger $\dG$ values are allowed.

Our phenomenological approach has been to lay out a range of quantities describing the strength of the gravitational field in and around the R16 Cepheids that may govern their screening behaviour. For given fifth force strength, the quantity most important for determining the resulting effect on $H_0$ is the fraction of Cepheids a given proxy leaves unscreened. In our formalism this is the fraction in ``weaker fields'' than the Solar System. It would therefore be helpful to formulate new models in which the onset of screening is governed by a variable maximally different between the MW and N4258 (and the LMC and M31), on the one hand, and the remainder of the sample on the other. The best property that we have discovered so far in this respect is $\Phi_{0.5}$ (the potential due to the immediate 0.5 Mpc environment), followed by the local dark matter density.

An assumption of all the present models is that the sources of screening contribute isotropically: this could be broken to allow for more complex dependence of degree of screening on environment, widening the parameter space in which to search for differences between the galaxies. 
One possibility is to construct models in which the threshold for screening is itself environment-dependent. This could happen for example in theories with several stable minima for the scalar, such as radiatively stable symmetrons \cite{Burrage:2016xzz} or supersymmetric models \cite{Brax:2012mq,Brax:2013yja}. Another is to incorporate a nontrivial geometry-dependence of the screening thresholds \cite{Burrage:2014daa,Bloomfield:2014zfa,Falck:2014jwa}. Investigating this possibility would require designing anisotropic environmental proxies such as the tidal field or higher moments of the potential, which may be aided by constrained simulations in modified gravity \cite{Shao}.

Finally, while we have opted simply to modify the R19 results according to the $G$-dependence of Cepheid properties, the analysis would ideally be performed by forward-modelling within a complete and self-consistent Bayesian likelihood framework. This would provide a more precise determination of the effect of fifth force parameters on the $H_0$ posterior and allow the degeneracies between those parameters and other degrees of freedom in the analysis to be mapped out. It would however require a  more thorough treatment of the $H_0$ likelihood function than is given here, in addition to more careful consideration of the metallicity dependence, outlier behaviour and potential nonlinearity of the PLR \cite{Ngeow, Becker, Kodric}. A joint analysis of both CMB and distance ladder data would enable posteriors on fifth force parameters to be derived by requiring a single $H_0$ value to account for both datasets, and quantify the overall increase in likelihood afforded by our models.

\section{Conclusion}
\label{sec:conc}

The $H_0$ tension is perhaps the most pressing problem in cosmology today. In recent years, with additional data and analyses the evidence appears to be mounting that the tension is due to new physics that invalidates direct comparison between low and high redshift probes of $H_0$ within $\Lambda$CDM. In this work, we set out to alleviate the $H_0$ tension by questioning the assumption that the physics of Cepheid stars is identical across the galaxies used to build the cosmic distance ladder.
In particular, we note that the period--luminosity relation of Cepheids is sensitive to the fifth forces that arise in most models beyond GR-based $\Lambda$CDM. The screening mechanisms these models are expected to employ may create a difference between the calibration and cosmological subsets of the distance ladder data, and neglect of this generically biases the local $H_0$ high. That is, the true $H_0$ is lower, closer to the value inferred from the CMB. 

We explore phenomenological screening models motivated by the range of mechanisms currently studied in the field. These include scalar--tensor theories of gravity as well as a new mechanism based on baryon--dark matter interactions that makes the effective strength of gravity a function of the local dark matter density $\rho_\text{DM}$ \cite{rho_DM}. Imposing the constraint that the  Cepheids used to calibrate the  period--luminosity relation are screened, we calculate the expected shift in the $H_0$ posterior as a function of fifth force strength relative to gravity, $\Delta G/G_{\rm N}$. This is done by modelling the change in both the Cepheid period and luminosity under a fifth force, then propagating this into the constraints on the host galaxy distances. These distances are combined with apparent magnitudes of SNe to calculate $H_0$. 

Our main results are shown in Figs.~\ref{fig:H0_proxies} and~\ref{fig:dG_discrep}. 
We find that models that unscreen Cepheids' cores (and hence raise their luminosities) lower the distance ladder result to at best within 2$-3\sigma$ of the \textit{Planck} value ($\sim$71--72 km s$^{-1}$ Mpc$^{-1}$) when $\Delta G/G_{\rm N} \approx 0.03-0.05$. Such a scenario is realised concretely in the baryon--dark matter interaction model. Models that only unscreen Cepheid envelopes (e.g. chameleons) require larger $\Delta G/G_{\rm N} \approx 0.3-0.4$ for this level of improvement, although they also evade constraints from MW stars more readily since stellar evolution is much less altered. We derive relatively strong (but possibly systematics-prone) bounds on fifth force strength in our models by comparing Cepheid and TRGB distance estimates, and weaker bounds from the universality of $M_\text{SN}$ and intrinsic PLR scatter, 
but conclude that there remains a viable region of screening parameter space that can reduce the Hubble tension to $2-3\sigma$. This requires Cepheid cores to be unscreened. Our models are no doubt further testable with data spanning various stages of stellar evolution, following earlier studies of distance indicators such as \cite{Jain}. We study fifth force effects on SNe in Appendix~\ref{Appendix:SN}.

A more complete resolution along these lines will require further work on the $H_0$ inference methodology. The full $H_0$ likelihood analysis should be repeated within specific candidate theories to better constrain fifth force parameters and map out their degeneracies with other degrees of freedom. A more thorough investigation of consistency tests internal to the distance ladder is also warranted, for example the scatter in individual galaxies' Cepheid period--magnitude relations. The effects of fifth forces on alternative $H_0$ inference methods, such as the TRGB-calibrated distance ladder and strong lensing time delays, need to be studied. Should the baryon--dark matter interaction model remain of interest, its predictions not only for astrophysical objects but also for cosmology must be investigated. Better than $2\sigma$ Cepheid compatibility with \textit{Planck} will require a qualitatively different screening mechanism than those currently under investigation, and theoretical developments may supply further insight in this regard. We conclude that there is much to explore on the implications of expansion rate measurements for local fundamental physics.

\bigskip

{\it Acknowledgements:} 
We are grateful for discussions with Koushik Sen, Radek Smolec and the wider MESA community for answering our various MESA-related questions. We thank Eric Baxter, Rachael Beaton, Cullen Blake, Ryosuke Hirai, Mike Jarvis, Justin Khoury, Lucas Macri, Helena Sanchez and Dan Scolnic for useful discussions. HD is supported by St John's College, Oxford, and acknowledges financial support from ERC Grant No. 693024 and the Beecroft Trust. BJ is supported in part by the US Department of Energy Grant No. DE-SC0007901 and by NASA ATP Grant No. NNH17ZDA001N. JS is supported by funds made available to the Center for Particle Cosmology by the University of Pennsylvania.

\bibliography{ref_H0}

\appendix

\section{Gravitational properties of Cepheid hosts and effects on $H_0$}
\label{sec:app1}

Tables~\ref{tab:galprops1} and~\ref{tab:galprops2} list the galaxy and halo properties of the R16 sample respectively, determined with the methods of Sec.~\ref{sec:models}. We use these as proxies for the degree of screening. Table~\ref{tab:H0} shows the effects of the models on $\hat{H}_0$, $\Delta \hat{H}_0$ and $\sigma_{H_0}$ for the maximum $\Delta G/G_\text{N}$ allowed at $5\sigma$ by the Cepheid vs TRGB distance test of Sec.~\ref{sec:tests} (Fig.~\ref{fig:trgb}), separately for each model and assuming the same unscreened fraction of galaxies in the R16 and TRGB samples.

\begin{table*}
  \begin{center}
  \small\addtolength{\tabcolsep}{-5pt}
    \begin{tabular}{|l|c|c|c|c|c|c|c|c|c|}
      \hline
      Name & RA/Dec (J2000 \degree) & $D$/Mpc & $\Delta D$/Mpc & $M_\text{SN}$ & $\Delta M_\text{SN}$ & $M_V$ & $\Delta M_V$ & $W_{20}/\text{km/s}$ & $\Delta W_{20}/\text{km/s}$\\ 
      \hline
    \rule{0pt}{3.5ex}
      MW & --- & 0 & 0 & --- & --- & -20.60 & 0.50 & 169 & 50\\
    \rule{0pt}{3.5ex}
      N4258 & 184.74 / 47.30 & 7.54 & 0.20 & --- & --- & -21.90 & 0.07 & 442 & 5\\
    \rule{0pt}{3.5ex}
      M101 & 210.80 / 54.35 & 6.71 & 0.14 & -19.39 & 0.13 & -21.30 & 0.20 & 194 & 5\\
    \rule{0pt}{3.5ex}
      N1015 & 39.55 / -1.32 & 31.58 & 1.18 & -19.05 & 0.15 & -20.89 & 0.24 & 188 & 10\\
    \rule{0pt}{3.5ex}
      N1309 & 50.53 / -15.40 & 31.96 & 0.81 & -19.33 & 0.13 & -20.80 & 0.20 & 161 & 6\\
    \rule{0pt}{3.5ex}
      N1365 & 53.40 / -36.14 & 18.26 & 0.48 & -19.39 & 0.14 & -23.00 & 0.20 & 404 & 5\\
    \rule{0pt}{3.5ex}
      N1448 & 56.13 / -44.64 & 18.29 & 0.38 & -19.11 & 0.13 & -21.99 & 0.23 & 414 & 7\\
    \rule{0pt}{3.5ex}
      N2442 & 144.10 / -69.53 & 20.05 & 0.49 & -19.24 & 0.15 & -21.90 & 0.20 & 514 & 21\\
    \rule{0pt}{3.5ex}
      N3021 & 147.74 / 33.55 & 31.59 & 1.31 & -19.54 & 0.15 & -21.20 & 0.20 & 291 & 8\\
    \rule{0pt}{3.5ex}
      N3370 & 161.77 / 17.27 & 25.97 & 0.59 & -19.16 & 0.13 & -21.20 & 0.20 & 287 & 12\\
    \rule{0pt}{3.5ex}
      N3447 & 163.36 / 16.78 & 24.08 & 0.48 & -19.21 & 0.13 & -17.90 & 0.20 & 116 & 32\\
    \rule{0pt}{3.5ex}
      N3972 & 178.94 / 55.32 & 20.77 & 0.67 & -19.10 & 0.14 & -20.68 & 0.18 & 266 & 11\\
    \rule{0pt}{3.5ex}
      N3982 & 179.12 / 55.13 & 22.25 & 0.71 & -19.51 & 0.13 & -21.14 & 0.10 & 232 & 7\\
    \rule{0pt}{3.5ex}
      N4038 & 180.47 / -18.87 & 18.11 & 0.93 & -19.06 & 0.16 & -21.84 & 0.19 & 294 & 8\\
    \rule{0pt}{3.5ex}
      N4424 & 186.80 / 9.42 & 16.44 & 2.21 & -19.53 & 0.31 & -19.20 & 0.20 & 95 & 5\\
    \rule{0pt}{3.5ex}
      N4536 & 188.61 / 2.19 & 15.18 & 0.37 & -19.29 & 0.14 & -22.97 & 0.20 & 353 & 6\\
    \rule{0pt}{3.5ex}
      N4639 & 190.72 / 13.26 & 20.25 & 0.66 & -19.11 & 0.14 & -21.32 & 0.07 & 303 & 7\\
    \rule{0pt}{3.5ex}
      N5584 & 215.60 / -0.39 & 22.76 & 0.48 & -19.09 & 0.12 & -21.00 & 0.20 & 215 & 5\\
    \rule{0pt}{3.5ex}
      N5917 & 230.39 / -7.38 & 28.35 & 1.33 & -19.26 & 0.15 & -18.70 & 0.20 & 237 & 6\\
    \rule{0pt}{3.5ex}
      N7250 & 334.57 / 40.56 & 19.94 & 0.72 & -19.20 & 0.14 & -19.60 & 0.20 & 203 & 10\\
    \rule{0pt}{3.5ex}
      U9391 & 218.65 / 59.34 & 38.35 & 1.11 & -19.45 & 0.13 & -18.50 & 0.20 & 140 & 15\\
    \hline
    \end{tabular}
  \caption{Properties of the period--luminosity relation anchors (MW \& N4258) and other extragalactic Cepheid hosts in the distance ladder analysis. $D$ and $M_\text{SN}$ are transcribed from R16, visual magnitude $M_V$ is obtained from the Nasa Extragalactic Database (NED) and HI linewidth $W_{20}$ is obtained from the Extragalactic Distance Database (EDD).}
  \label{tab:galprops1}
  \end{center}
\end{table*}

\begin{table*}
  \begin{center}
  \small\addtolength{\tabcolsep}{-5pt}
    \begin{tabular}{|l|c|c|c|c|c|c|c|c|}
      \hline
      Name & $\log(R/\text{kpc})$ & $\Delta \log(R/\text{kpc})$ & $\log(c)$ & $\Delta \log(c)$ & $\log(V/\text{km/s})$ & $\Delta \log(V/\text{km/s})$ & $\log(M/M_\odot)$ & $\Delta \log(M/M_\odot)$\\ 
      \hline
    \rule{0pt}{3.5ex}
      MW & 2.37 & 0.12 & 1.07 & 0.27 & 2.24 & 0.12 & 11.94 & 0.37\\
    \rule{0pt}{3.5ex}
      N4258 & 2.75 & 0.13 & 0.93 & 0.2 & 2.45 & 0.13 & 13.01 & 0.39\\
    \rule{0pt}{3.5ex}
      M101 & 2.61 & 0.14 & 0.99 & 0.22 & 2.31 & 0.14 & 12.59 & 0.43\\
    \rule{0pt}{3.5ex}
      N1015 & 2.54 & 0.13 & 1.01 & 0.23 & 2.24 & 0.13 & 12.39 & 0.38\\
    \rule{0pt}{3.5ex}
      N1309 & 2.51 & 0.13 & 1.03 & 0.23 & 2.21 & 0.13 & 12.31 & 0.39\\
    \rule{0pt}{3.5ex}
      N1365 & 3.02 & 0.15 & 0.83 & 0.16 & 2.72 & 0.15 & 13.83 & 0.44\\
    \rule{0pt}{3.5ex}
      N1448 & 2.76 & 0.14 & 0.93 & 0.20 & 2.46 & 0.14 & 13.04 & 0.43\\
    \rule{0pt}{3.5ex}
      N2442 & 2.75 & 0.14 & 0.93 & 0.19 & 2.45 & 0.14 & 13.00 & 0.42\\
    \rule{0pt}{3.5ex}
      N3021 & 2.59 & 0.13 & 1.00 & 0.23 & 2.29 & 0.13 & 12.55 & 0.39\\
    \rule{0pt}{3.5ex}
      N3370 & 2.59 & 0.13 & 1.00 & 0.22 & 2.29 & 0.13 & 12.52 & 0.39\\
    \rule{0pt}{3.5ex}
      N3447 & 2.25 & 0.13 & 1.10 & 0.28 & 1.95 & 0.13 & 11.52 & 0.38\\
    \rule{0pt}{3.5ex}
      N3972 & 2.51 & 0.11 & 1.01 & 0.23 & 2.21 & 0.11 & 12.30 & 0.34\\
    \rule{0pt}{3.5ex}
      N3982 & 2.59 & 0.12 & 0.99 & 0.23 & 2.29 & 0.12 & 12.53 & 0.35\\
    \rule{0pt}{3.5ex}
      N4038 & 2.72 & 0.15 & 0.97 & 0.19 & 2.42 & 0.15 & 12.93 & 0.44\\
    \rule{0pt}{3.5ex}
      N4424 & 2.34 & 0.13 & 1.07 & 0.26 & 2.04 & 0.13 & 11.79 & 0.38\\
    \rule{0pt}{3.5ex}
      N4536 & 3.00 & 0.16 & 0.86 & 0.17 & 2.70 & 0.16 & 13.77 & 0.48\\
    \rule{0pt}{3.5ex}
      N4639 & 2.61 & 0.12 & 1.00 & 0.22 & 2.31 & 0.12 & 12.60 & 0.37\\
    \rule{0pt}{3.5ex}
      N5584 & 2.56 & 0.12 & 1.00 & 0.24 & 2.26 & 0.12 & 12.43 & 0.37\\
    \rule{0pt}{3.5ex}
      N5917 & 2.30 & 0.13 & 1.08 & 0.29 & 2.00 & 0.13 & 11.66 & 0.38\\
    \rule{0pt}{3.5ex}
      N7250 & 2.38 & 0.12 & 1.06 & 0.26 & 2.08 & 0.12 & 11.91 & 0.35\\
    \rule{0pt}{3.5ex}
      U9391 & 2.29 & 0.12 & 1.09 & 0.28 & 1.99 & 0.12 & 11.62 & 0.37\\
      \hline
    \end{tabular}
  \caption{Halo properties of the galaxy sample, estimated by abundance matching to $M_V$ (Sec.~\ref{sec:models}). Size $R$, rotation velocity $V$ and enclosed mass $M$ are evaluated at the virial radius.}
  \label{tab:galprops2}
  \end{center}
\end{table*}

\begin{table*}[]
  \begin{center}
  \small\addtolength{\tabcolsep}{-5pt}
    \begin{tabular}{|c|c|c|c|c|c|c|}
      \hline
      Proxy & Unscr frac & Screening properties & $\dG$ & $\hat{H}_0$ / km s$^{-1}$ Mpc$^{-1}$ & $\Delta \hat{H}_0$ / km s$^{-1}$ Mpc$^{-1}$ & $\sigma_{H_0}$\\ 
      \hline      
\rule{0pt}{3.5ex}
      --- & --- & $0\: / \:0\: / \:0\: / \:0$ & 0 & 74.0 & 1.4 & 4.4\\
      \hline     
\rule{0pt}{3.5ex}
      $\Phi_{0.5}$ & 0.85 & $1\: / \:0\: / \:0\: / \:0$ & 0.09 & 72.4 & 1.4 & 3.3\\
      \hline
\rule{0pt}{3.5ex}
      $\Phi_{5}$  & 0.08 & $1\: / \:0\: / \:0\: / \:0$ & 0.22 & 73.4 & 1.5 & 3.9\\
      \hline
\rule{0pt}{3.5ex}
      $\Phi_{50}$ & 0.30 & $1\: / \:0\: / \:0\: / \:0$ & 0.14 & 72.8 & 1.5 & 3.5\\
      \hline
\rule{0pt}{3.5ex}
      $a_{0.5}$ & 0.31 & 
      \begin{tabular}{m{2.12cm}}\parbox[c]{2.12cm}{$1\: / \:0\: / \:0\: / \:0$\\$1\: / \:1\: / \:0\: / \:0$}\end{tabular} & \begin{tabular}{m{1.1cm}}\parbox[c]{1.1cm}{0.14\\0.04}\end{tabular} & \begin{tabular}{m{1.1cm}}\parbox[c]{1.1cm}{72.8\\71.9}\end{tabular} & \begin{tabular}{m{0.97cm}}\parbox[c]{0.97cm}{1.5\\1.6}\end{tabular} & \begin{tabular}{m{1.25cm}}\parbox[c]{1.25cm}{\hspace{0.85mm}3.5\\2.6}\end{tabular}\\
      \hline
\rule{0pt}{3.5ex}
      $a_5$ & 0.31 & \begin{tabular}{m{2.12cm}}\parbox[c]{2.12cm}{$1\: / \:0\: / \:0\: / \:0$\\$1\: / \:1\: / \:0\: / \:0$}\end{tabular} & \begin{tabular}{m{1.1cm}}\parbox[c]{1.1cm}{0.14\\0.04}\end{tabular} & \begin{tabular}{m{1.1cm}}\parbox[c]{1.1cm}{73.1\\72.2}\end{tabular} & \begin{tabular}{m{0.97cm}}\parbox[c]{0.97cm}{1.5\\1.8}\end{tabular} & \begin{tabular}{m{0.97cm}}\parbox[c]{0.97cm}{3.6\\2.6}\end{tabular}\\
      \hline
\rule{0pt}{3.5ex}
      $a_{50}$ & 0.35 & \begin{tabular}{m{2.12cm}}\parbox[c]{2.12cm}{$1\: / \:0\: / \:0\: / \:0$\\$1\: / \:1\: / \:0\: / \:0$}\end{tabular} & \begin{tabular}{m{1.1cm}}\parbox[c]{1.1cm}{0.13\\0.04}\end{tabular} & \begin{tabular}{m{1.1cm}}\parbox[c]{1.1cm}{72.6\\71.4}\end{tabular} & \begin{tabular}{m{0.97cm}}\parbox[c]{0.97cm}{1.4\\1.6}\end{tabular} & \begin{tabular}{m{1.25cm}}\parbox[c]{1.25cm}{\hspace{0.85mm}3.5\\2.5}\end{tabular}\\
      \hline
\rule{0pt}{3.5ex}
      $K_{0.5}$ & 0.39 & \begin{tabular}{m{2.12cm}}\parbox[c]{2.12cm}{$1\: / \:0\: / \:0\: / \:0$\\$1\: / \:1\: / \:0\: / \:0$}\end{tabular} & \begin{tabular}{m{1.1cm}}\parbox[c]{1.1cm}{0.13\\0.03}\end{tabular} & \begin{tabular}{m{1.1cm}}\parbox[c]{1.1cm}{72.8\\71.4}\end{tabular} & \begin{tabular}{m{0.97cm}}\parbox[c]{0.97cm}{1.4\\1.5}\end{tabular} & \begin{tabular}{m{1.25cm}}\parbox[c]{1.25cm}{\hspace{0.85mm}3.6\\2.5}\end{tabular}\\
      \hline
\rule{0pt}{3.5ex}
      $K_{5}$ & 0.30 & \begin{tabular}{m{2.12cm}}\parbox[c]{2.12cm}{$1\: / \:0\: / \:0\: / \:0$\\$1\: / \:1\: / \:0\: / \:0$}\end{tabular} & \begin{tabular}{m{1.1cm}}\parbox[c]{1.1cm}{0.14\\0.04}\end{tabular} & \begin{tabular}{m{1.1cm}}\parbox[c]{1.1cm}{72.6\\71.8}\end{tabular} & \begin{tabular}{m{0.97cm}}\parbox[c]{0.97cm}{1.4\\1.5}\end{tabular} & \begin{tabular}{m{1.25cm}}\parbox[c]{1.25cm}{\hspace{0.85mm}3.4\\2.7}\end{tabular}\\
      \hline
\rule{0pt}{3.5ex}
      $K_{50}$ & 0.26 & \begin{tabular}{m{2.12cm}}\parbox[c]{2.12cm}{$1\: / \:0\: / \:0\: / \:0$\\$1\: / \:1\: / \:0\: / \:0$}\end{tabular} & \begin{tabular}{m{1.1cm}}\parbox[c]{1.1cm}{0.14\\0.04}\end{tabular} & \begin{tabular}{m{1.1cm}}\parbox[c]{1.1cm}{72.9\\71.9}\end{tabular} & \begin{tabular}{m{0.97cm}}\parbox[c]{0.97cm}{1.5\\1.6}\end{tabular} & \begin{tabular}{m{0.97cm}}\parbox[c]{0.97cm}{3.6\\2.7}\end{tabular}\\
      \hline
\rule{0pt}{3.5ex}
      $L_\text{gal}$ & 0.12 & \begin{tabular}{m{2.12cm}}\parbox[c]{2.12cm}{$1\: / \:0\: / \:0\: / \:0$\\$1\: / \:1\: / \:0\: / \:0$}\end{tabular} & \begin{tabular}{m{1.1cm}}\parbox[c]{1.1cm}{0.19\\0.06}\end{tabular} & \begin{tabular}{m{1.1cm}}\parbox[c]{1.1cm}{73.5\\72.0}\end{tabular} & \begin{tabular}{m{0.97cm}}\parbox[c]{0.97cm}{1.4\\1.4}\end{tabular} & \begin{tabular}{m{0.97cm}}\parbox[c]{0.97cm}{4.0\\3.1}\end{tabular}\\
      \hline  
\rule{0pt}{3.5ex}
      $W_{20}$ & 0.20 & \begin{tabular}{m{2.12cm}}\parbox[c]{2.12cm}{$1\: / \:0\: / \:0\: / \:0$\\$1\: / \:1\: / \:0\: / \:0$}\end{tabular} & \begin{tabular}{m{1.1cm}}\parbox[c]{1.1cm}{0.16\\0.05}\end{tabular} & \begin{tabular}{m{1.1cm}}\parbox[c]{1.1cm}{73.5\\72.9}\end{tabular} & \begin{tabular}{m{0.97cm}}\parbox[c]{0.97cm}{1.5\\1.9}\end{tabular} & \begin{tabular}{m{0.97cm}}\parbox[c]{0.97cm}{3.8\\2.8}\end{tabular}\\
      \hline
\rule{0pt}{3.5ex}
      $M_\text{vir}$ & 0.37 & \begin{tabular}{m{2.12cm}}\parbox[c]{2.12cm}{$1\: / \:0\: / \:0\: / \:0$\\$1\: / \:1\: / \:0\: / \:0$}\end{tabular} & \begin{tabular}{m{1.1cm}}\parbox[c]{1.1cm}{0.13\\0.04}\end{tabular} & \begin{tabular}{m{1.1cm}}\parbox[c]{1.1cm}{72.9\\71.5}\end{tabular} & \begin{tabular}{m{0.97cm}}\parbox[c]{0.97cm}{1.5\\1.8}\end{tabular} & \begin{tabular}{m{1.25cm}}\parbox[c]{1.25cm}{\hspace{0.85mm}3.5\\2.2}\end{tabular}\\
      \hline
\rule{0pt}{3.5ex}
      $\rho_\text{DM}$ & 0.45 & \begin{tabular}{m{2.12cm}}\parbox[c]{2.12cm}{$1\: / \:1\: / \:0\: / \:0$\\$1\: / \:1\: / \:1\: / \:0$}\end{tabular} & \begin{tabular}{m{1.1cm}}\parbox[c]{1.1cm}{0.03\\0.03}\end{tabular} & \begin{tabular}{m{1.1cm}}\parbox[c]{1.1cm}{71.8\\72.4}\end{tabular} & \begin{tabular}{m{0.97cm}}\parbox[c]{0.97cm}{1.5\\1.5}\end{tabular} & \begin{tabular}{m{1.25cm}}\parbox[c]{1.25cm}{2.8\\3.3}\end{tabular}\\
      \hline
\rule{0pt}{3.5ex}
      E. U. & 1 & $0\: / \:0\: / \:0\: / \:1$ & 0.05 & 71.5 & 1.4 & 2.8\\
      \hline
    \end{tabular}
  \caption{The effect of various screening proxies on the value of $H_0$ inferred from the local distance ladder, assuming a $\Delta G/G_\text{N}$ at the $5\sigma$ limit of that allowed by the Cepheid--TRGB distance test, separately for each proxy. The first row indicates the R19 result. In the third column, ``1'' indicates unscreened (feels fifth force) and ``0'' screened for Cepheid envelope/Cepheid core/Calibration SNe/Cosmological SNe. As described in the text, the threshold value for screening under a given proxy is given by the MW's value for that proxy. $\sigma_{H_0}$ is the discrepancy of the resulting $H_0$ constraint with the CMB (Eq.~\eqref{eq:sig_H0}). ``E.~U.'' denotes the ``early unscreening'' model (Appendix~\ref{Appendix:SN}) in which cosmological SNe are fully unscreened and calibration SNe fully screened, without any effect on Cepheids. In this case we consider $\dG=0.05$. Our best models, in which Cepheid cores are unscreened, achieve $\sim$2$-3\sigma$ consistency with \textit{Planck}.}
  \label{tab:H0}
  \end{center}
\end{table*}

\section{Unscreening supernovae}
\label{Appendix:SN}

\subsection{Fifth force effects on SNe}
\label{sec:F5_SNe}

Under a fifth force, the Chandrasekhar mass of white dwarfs, the progenitors of SNe, is reduced, resulting in less fuel available for the thermonuclear explosion. The simplest model for the SN luminosity is $L \propto M_\text{Ch} \propto G^{-3/2}$ \cite{SN_G1, SN_G2}, although this neglects the variation in the mass of Ni$^{56}$, the primary isotope responsible for the luminosity, as well as the standardisation of $L$ with the width of the light curve. Including both of these effects in a full semi-analytic model reverses the trend of the $L$-$G$ relation \cite{Wright}. We model this as
\begin{equation}\label{eq:C}
L_\text{SN} \propto G^C,
\end{equation}
which implies a fifth force induced change of
\begin{equation}\label{eq:SNe_mag}
\Delta M_{\rm SN} = - 2.5 \: C \: \log(1 + \Delta G_{\rm SN}/G_\text{N}),
\end{equation}
for $G_{\rm SN}$ the effective Newton's constant of the SN in question. We set $C = 1.46$ from a linear fit to \cite{Wright} fig. 7 (left). This modification is however not present when SNe are screened (as expected in chameleon and similar models, since the potential is at least an order of magnitude higher in white dwarfs than Cepheids), which may be modelled by setting $C=0$.

\subsection{SN effects on $H_0$}

Including the above effects as well as the Cepheid distance modification of Eq.~\eqref{eq:d_avg}, the $H_0$ value implied by a given galaxy is modified to
\begin{align}\label{eq:H0_22}
\bar{H}_{0,i} &= 10^{m_\text{SN,i}/5 + 6 + a_B} \: \frac{\text{pc}}{\bar{D}_i} \: \left(\frac{G_{\rm SN,i}}{G_\text{N}}\right)^{-C/2} \: \text{km s}^{-1} \text{Mpc}^{-1}\\ \nonumber
&= \frac{H_0^\text{R19}}{K_i} \: \left(\frac{G_{\rm SN,i}}{G_\text{N}}\right)^{-C/2},
\end{align}
so that the total $H_0$ including all galaxies is given by
\begin{equation}\label{eq:H0_44}
\bar{H}_0 = H_0^\text{R19} \: \frac{\sum_i K_i^{-1} \left(\frac{G_{\text{SN},i}}{G_{\rm N}}\right)^{-C/2}{\Delta \bar{H}_{0,i}^{-2}}}{\sum_i {\Delta \bar{H}_{0,i}^{-2}}}\:,
\end{equation}
which is the equivalent of Eq. \eqref{eq:H0_4} including SNe unscreening.

Because a fifth force acts to reduce SN magnitude, a decrease in the inferred $H_0$ comes about when the calibration SNe are more screened on average than the cosmological SNe, with Eqs.~\eqref{eq:H0_22}--\eqref{eq:H0_44} formally requiring that the former are fully screened and the latter fully unscreened. However, as described in Sec.~\ref{sec:method} we do not model the cosmological SNe explicitly but rather adopt the magnitude--redshift relation already fitted in R16. Hence we cannot explicitly model differences in screening between the two samples, justifying our assumption in the main text that in the absence of further information the calibration and cosmological samples are likely to feel the fifth force on average in the same way, resulting in no further change to $H_0$ (effectively $C=0$). For the $\rho_\text{DM}$ model, this could be further justified by noting that the average density of halos is roughly constant over cosmic time.

To explore the potential impact of the SN luminosity shift on $H_0$ we also consider two models in which the calibration and cosmological SNe are differently screened. The first uses the $\rho_\text{DM}$ proxy and assumes that the cosmological SNe are fully screened while the calibration SNe are unscreened, which effectively corresponds to $C=-1.46$ in Eq.~\eqref{eq:H0_44}. This is less likely a priori than the fiducial case, but \emph{increases} $H_0$ and gives an indication of the relative importance of the Cepheid and SN effects.
We assume in this case that a SN is as unscreened on average as the Cepheids in the galaxy that hosts it. Since the precise locations of the SNe within their host galaxies are unknown, we use the average $G_{ij}$ across the galaxies' Cepheids as a proxy for $G_\text{SN}$: $G_{\text{SN},i} = \sum_j G_{ij}/N_i$ for $N_i$ Cepheids in galaxy $i$. The result is shown in the $\rho_\text{DM}$ row of Table \ref{tab:H0}, where we see that this effect is subdominant to the shift in Cepheid distance and hence that the $H_0$ tension may still be ameliorated with $\Delta G/G_\text{N} \approx 0.05$.

The second scenario with nonzero $C$ that we consider assumes that all SNe feel the full fifth force, and that screening occurs at low redshift $z \lesssim 0.1$. In this case all Cepheids and the calibration sample of SNe are screened, while the cosmological SNe are unscreened (effectively $C=1.46$, $\bar{D}_i = D_i$). We dub this model \emph{early unscreening} (E.~U.) and show the result in the final row of Table \ref{tab:H0}. $\Delta G/G_\text{N} \approx 0.1$ would be required for $H_0$ to reach the \textit{Planck} value in this scenario.

Early unscreening is harder to realise theoretically than the other models because screening mechanisms are designed to activate at high density, implying that screening tends to be more efficient at higher redshift. One possibility is to engineer the cosmological dynamics such that the threshold for screening decreases at earlier times. A transition redshift of $z\simeq0.1$, which roughly separates the calibration and cosmological SNe, could be achieved in a chameleon-like model if the cosmological field, which sets the screening threshold, is pinned to a large value at early times but begins to roll and track the minimum of the effective potential around the transition redshift so that screening only occurs later on. Such a model would likely require the force mediator to couple preferentially to carbon and oxygen, since a significant coupling to hydrogen and helium would require the field to track its minimum throughout cosmic history. If this is not the case then the masses of these elements would have changed significantly over cosmic history \cite{Brax:2004qh}, and the distance to the surface of last scattering would also be altered.





\end{document}